\documentclass[a4paper,fleqn,usenatbib]{mnras}

\usepackage{mathptmx}

\usepackage[T1]{fontenc}
\usepackage{ae,aecompl}

\usepackage{graphicx}	
\usepackage{amsmath}	
\usepackage{amssymb}	
\usepackage{setspace}

\newcommand{\pcm}{cm$^{-2}$}
\newcommand{\xmm}{\textit{XMM-Newton}}
\newcommand{\nustar}{\textit{NuSTAR}}

\newcommand{\delcstat}{$\Delta$C-stat}
\newcommand{\chandra}{\textit{Chandra}}

\title[Ionised Emission and Absorption in ULXs]{Ionised Emission and Absorption in a Large Sample of Ultraluminous X-ray Sources}

\author[P Kosec et al.]{P. Kosec$^{1,2}$\thanks{E-mail: pkosec@mit.edu}, 
C. Pinto$^{3,4}$,
C. S. Reynolds$^{1}$,
M. Guainazzi$^{4}$,
E. Kara$^{2}$,
D. J. Walton$^{1}$,\newauthor
A. C. Fabian$^{1}$,
M. L. Parker$^{1,5}$,
and I. Valtchanov$^{6}$
\\
$^{1}$Institute of Astronomy, Madingley Road, CB3 0HA Cambridge, UK \\
$^{2}$MIT Kavli Institute for Astrophysics and Space Research, Cambridge, MA 02139, USA\\
$^{3}$INAF - IASF Palermo, Via U. La Malfa 153, I-90146 Palermo, Italy\\
$^{4}$ESTEC/ESA, Keplerlaan 1, 2201AZ Noordwijk, The Netherlands\\
$^{5}$European Space Astronomy Centre (ESAC), Science Operations Department, 28692 Villanueva de la Ca\~nada, Madrid, Spain\\
$^{6}$Telespazio Vega UK for ESA, European Space Astronomy Centre, Operations Department, 28691 Villanueva de la Ca\~nada, Spain\\
}

\date{Accepted 2021 September 29. Received 2021 September 3; in original form 2021 May 21.}

\pubyear{2021}

\begin{document}
\label{firstpage}
\pagerange{\pageref{firstpage}--\pageref{lastpage}}
\maketitle

\begin{abstract}

Most Ultraluminous X-ray sources (ULXs) are thought to be powered by super-Eddington accretion onto stellar-mass compact objects. Accretors in this extreme regime are naturally expected to ionise copious amounts of plasma in their vicinity and launch powerful radiation-driven outflows from their discs.
High spectral resolution X-ray observations (with RGS gratings onboard XMM-Newton) of a few ULXs with the best datasets indeed found complex line spectra and confirmed such extreme ($0.1-0.3$c) winds. However, a search for plasma signatures in a large ULX sample with a rigorous technique has never been performed, thereby preventing us from understanding their statistical properties such as the rate of occurrence, to constrain the outflow geometry and its duty cycle.
We developed a fast method for automated line detection in X-ray spectra and applied it to the full RGS ULX archive, rigorously quantifying the statistical significance of any candidate lines. Collecting the 135 most significant features detected in 89 observations of 19 objects, we created the first catalogue of spectral lines detected in soft X-ray ULX spectra.
We found that the detected emission lines are concentrated around known rest-frame elemental transitions and thus originate from low-velocity material. The absorption lines instead avoid these transitions, suggesting they were imprinted by blueshifted outflows. Such winds therefore appear common among the ULX population. Additionally, we found that spectrally hard ULXs show fewer line detections than soft ULXs, indicating some difference in their accretion geometry and orientation, possibly causing over-ionisation of plasma by the harder spectral energy distributions of harder ULXs.

\end{abstract}




\section{Introduction}

\subsection{Ultraluminous X-ray sources}

Ultraluminous X-ray sources \citep{Kaaret+17} are non-nuclear point sources in nearby galaxies with an isotropic luminosity exceeding 10$^{39}$ erg/s, the Eddington luminosity of a standard 10 M$_{\odot}$ stellar-mass black hole. The nature of ULXs was an open question for decades, with two scenarios being that they are either powered by sub-Eddington accretion onto the elusive intermediate mass black holes \citep{Miller+03, Farrell+09} or by super-Eddington accretion onto stellar-mass compact objects \citep{King+01, Poutanen+07, Gladstone+09, King+09}.

Now, we know that a majority of these objects is powered by super-Eddington accretion. A crucial piece of evidence for this hypothesis was the discovery that at least some ULXs are pulsating and thus are unambiguously powered by neutron stars \citep{Bachetti+14, Furst+16, Israel+17a, Israel+17b, Carpano+18, Sathyaprakash+19, Rodriguez+20}. Furthermore, ULX X-ray spectra are substantially different from the typical spectra of sub-Eddington accreting systems because of a high-energy roll-over at energies less than 10 keV \citep{Bachetti+13, Walton+14}.

Moreover, powerful outflows with velocities of up to $20-30$\% of the speed of light were discovered in several ULXs through detection of highly blueshifted atomic features seen in absorption \citep{Pinto+16, Walton+16, Pinto+17, Kosec+18a}, including evidence for an outflow in a neutron star powered ULX \citep{Kosec+18b}. Radiation-driven disc winds are naturally expected to arise in super-Eddington accretion \citep{Shakura+73, Poutanen+07} and are therefore an important piece of evidence towards our understanding of the nature of ULXs.

\subsection{Automated spectral search methods}
\label{intro_oldmethods}

Any ionised plasma, which may or may not be part of the outflowing wind, will imprint spectral lines on the continuum X-ray spectrum of the ULX. If the plasma lies along our line of sight towards the X-ray source, it will absorb X-ray photons, producing absorption lines. Otherwise, it will produce emission lines. If the plasma is in motion with respect to us, the observed line wavelengths will be Doppler-shifted from their rest-frame wavelengths.

Considering that any detected spectral feature could then be Doppler-shifted, it is not always straightforward to identify it, in particular if the expected shifts can be large (which is the case here). Similarly, it is not easy to assign a statistical significance to any line detection - in other words, it is not straightforward to calculate the probability that the detected line originates from noise rather than from optically-thin plasma near the ULX. Given the low X-ray fluxes of ULXs (due to their Megaparsec distances), these features can often appear to be just on the limit of significance. Thus it is important to estimate their significance accurately.

This is a complex problem which must be approached systematically. An automated routine is necessary in order to search the source X-ray spectrum and locate any possible spectral lines. Then the algorithm needs to calculate the statistical probability that each of the tentatively detected features is real and does not originate from noise.

Such methods have recently been developed and successfully applied to detect outflows in a few ULXs \citep{Pinto+16, Pinto+17} and in AGN \citep{Cappi+09, Tombesi+10b, Kosec+20b}. These methods directly fit the X-ray spectra in spectral fitting packages such as \textsc{spex} \citep{Kaastra+96} and \textsc{xspec} \citep{Arnaud+96} in an automated fashion. The simpler versions of the routines scan the spectra with Gaussian lines \citep{Kosec+18a} and thus are used to detect absorption and emission lines. The more complex versions \citep{Kosec+18b, Pinto+20} search the spectra with large grids of physical models and are therefore able to identify and fit many absorption/emission lines simultaneously (accounting for their Doppler shifts) and infer the plasma physical properties.

These methods are very powerful in locating the best-fitting spectral features (lines or absorbers/emitters), however they become computationally expensive when one needs to assess the statistical significance of the detections. A search of a spectrum for spectral features will often locate many spurious features which might appear significant. This is simply because a large parameter space has been searched, increasing the chances that it contains a strong feature originating from pure noise \citep{Vaughan+08}. This is called the look-elsewhere effect. To account for such effect, one must repeat the search many times on simulated datasets with statistics comparable with the real data, but containing just Poisson noise superimposed on the featureless broadband continuum of the X-ray source. The false positive rate of any detected feature is then the fraction of simulated datasets containing fake features equal to or stronger than the one detected in real data.

As a result, to test for reasonable significances ($\sim3\sigma$) one has to perform a few 1000s Monte Carlo searches on simulated datasets, which can be very computationally expensive ($>$1000 computer hours). This kind of analysis is therefore only possible on a very small number of sources.

Here we make an important point regarding the line significances. All of the above only applies to completely random and unexpected features, i.e. lines that occur at wavelengths which do not belong to known and expected rest-frame elemental transitions. If a line is detected at a known transition and such transition is a plausible origin for the feature (no Doppler shift expected or the Doppler shift is known), the possible parameter space for its identification collapses and the line significance can be approximately \citep[but not exactly, as shown by][]{Protassov+02} determined directly from spectral fitting without the need for Monte Carlo simulations. In this case, the required line strength for detection significance is much lower. For a line with fixed wavelength (to the elemental transition rest-frame) and fixed width, this is a problem with a single extra free parameter (the line normalization), and thus a fit improvement of \delcstat\ $=9$ or $\Delta\chi^2=9$ results in 3$\sigma$ detection significance. Given the extreme conditions near the ULX, the number of plausible elemental transitions in the soft X-ray band is in fact quite small. Due to the high temperatures involved, the ULX spectra will likely only show a handful of K and L shell transitions of a few abundant elements such as Mg, Ne, Fe, O and N.

\subsection{This work}

In this work, we study all ULXs with high enough quality X-ray data, and search their high-resolution spectra for any emission or absorption lines, while quantifying the true statistical significance of any features detected. All the detected features are included in a single catalogue, thus creating the first catalogue of spectral line detections in ULXs. The catalogue will allow us to make the first statistical comparison of spectral lines observed in ULXs, important to obtain model-independent diagnostics on the line significance, rate of occurrence and nature. It will also be crucial for future observations and more sensitive missions such as \textit{XRISM} and \textit{Athena}, by highlighting promising ULXs to observe in future observational campaigns.

This work builds upon our first systematic search for spectral features in ULXs \citep[described in][]{Kosec+18a}, where a smaller sample (10 sources) was studied using the traditional Gaussian search method. Most of the detected significances were quantified only tentatively. We performed the fully rigorous search with simulations on the four most promising sources.

Here the sample is expanded to include all suitable ULX datasets (19 sources in total). As it is prohibitively expensive to search the full sample with the current automated methods (assessing the detection significances rigorously), we have developed a new, fast search method, employing cross-correlation to search X-ray spectra for spectral features. In this paper we apply the method to scan ULX spectra for Gaussian lines. However, the method can in principle be used to search the spectrum of any astrophysical source from any instrument for any spectral feature of interest.

Section \ref{ulxsample} contains the description of the ULX sample we created for this study. Section \ref{method} describes the newly developed cross-correlation method step-by-step. Section \ref{results} shows its performance and an example analysis of a well-studied ULX (NGC 1313 X-1). This section also contains the statistical results from the full object sample. We discuss and interpret these results in Section \ref{discussion}. Finally, Section \ref{conclusions} lists our conclusions. This is followed by appendices: Appendix \ref{cataloguestructure} explains the structure of the line catalogue, while Appendix \ref{methodappendix} gives a detailed description of each step of the cross-correlation method and Appendix \ref{ulxexpostats} gives more information about the individual sources studied and their broadband continuum fitting.

Throughout this paper, we use Cash statistics \citep[C-stat,][]{Cash+79} for spectral fitting and all uncertainties are stated at $1\sigma$ level.

\section{The ULX sample}
\label{ulxsample}

Due to their extragalactic distances, ULXs are too faint to provide high quality datasets with the High Energy Transmission Gratings \citep[HETG,][]{Canizares+05} onboard the \chandra\ observatory, unless very long exposures are used. Increasing the necessary exposure time increases the probability that the line features shift or disappear, considering that they have been observed to vary in time in some ULXs \citep[e.g.][]{Kosec+18b, Pinto+20}, as expected for X-ray binary winds.

The other high-spectral resolution X-ray instrument in current use is the Reflection Grating Spectrometer \citep[RGS,][]{denHerder+01} onboard \xmm\ \citep{Jansen+01}. Its collecting area is significantly higher than that of HETG but its energy band is narrower ($0.35-1.8$ keV RGS bandpass versus $0.4-10$ keV HETG). However, this is the energy band which contains nitrogen, oxygen, neon, magnesium and iron transitions and is thus of great importance \citep{Kaastra+08}. These elements normally provide the strongest lines in X-ray binary and active galactic nuclei spectra unless the elements are too ionised. The RGS is therefore the main instrument of this study.

The imaging CCD-based and silicon-drift X-ray instruments such as EPIC \citep{Struder+01, Turner+01} onboard \xmm, ACIS onboard \chandra\ \citep{Weisskopf+00}, \textit{NICER} \citep{Gendreau+16} and \textit{eROSITA} \citep{Predehl+21} offer a spectral resolution of roughly 100 eV. This is too poor to resolve individual emission or absorption lines unless they are isolated, particularly in the soft X-ray band ($0.3-2$ keV). We therefore do not search data from X-ray CCD-based instruments for narrow spectral features in this work. Nevertheless, we use \xmm\ EPIC data to constrain the broadband continua of ULXs in the $0.3-10$ keV energy range.

We select all ULX observations with good enough quality RGS data for our sample. The criteria for a good quality dataset have previously been defined by \citet{Kosec+18a}, but we summarise them again here for clarity. The RGS instrument must be pointed directly at the ULX position, or with a small offset (less than 1 arcmin) due to the small RGS field of view. The RGS source region should not be contaminated by any other bright X-ray sources. Finally, the exposure should be such that the resulting spectrum (RGS 1 + RGS 2) contains at least 1000 source counts. In case the total count number is lower, it is possible to stack multiple observations of the source with the of risk washing out any transient spectral features.

In addition to ULXs with persistent super-Eddington activity, we also select two Magellanic Cloud X-ray pulsars with luminosities (temporarily) exceeding their Eddington limits ($L \sim 10^{38}$ erg/s and higher): SMC X-3 and RX J0209.6-7427. Given that at least a fraction of ULXs are powered by neutron stars, there could be many similarities between ULXs and these transient super-Eddington pulsars (even though the latter do not always reach above the luminosities of 10$^{39}$ erg/s). We also considered including the Galactic pulsar Swift J0243.6+6124 (which temporarily reached ULX luminosity levels in 2017) in the sample. However, the only pointed \xmm\ observation occurred when the source was significantly below the Eddington limit and its \chandra\ HETG observation appears to be plagued by instrumental systematics \citep{Eijnden+19}. Therefore we do not study Swift J0243.6+6124.

The final sample contains 17 ULXs and 2 super-Eddington pulsars, and is shown in Table \ref{ulxtable}. The sample contains all the suitable \xmm\ observations of ULXs available as of June 2020. The table lists the source name and the RGS observations of each object used in this work. For several sources, we used multiple approaches to search for any possible line features in their spectra. For example, we used a single high quality observation but at the same time we also tested the highest-statistics spectrum obtained by stacking the RGS spectra from all the available observations. These different approaches have designated names to identify them. Appendix \ref{ulxexpostats} contains further details about each approach for every object, listing the clean RGS exposures and the total net counts in the RGS spectra. It also lists the calculated hardness ratio of each source as well as the number of detected lines in its spectrum.

\begin{table*}
	\centering
	\caption{Sources used to create the ULX catalogue. The second column lists the name of the approach based on the observations used, the individual observations are shown in the third column. Multiple observations in the third column indicate that we searched in the stacked dataset from all of the observations listed. Further details of the individual approaches are listed in Appendix \ref{ulxexpostats}.} 
	\vspace{0.3cm}
	\label{ulxtable}
	\begin{tabular}{ccc}
		\hline
		Object name&Approach&Observations\\
		\hline

		Circinus ULX-5&&0701981001\\
		&&0824450301\\		
		
		Holmberg II X-1	&&0200470101\\
		&Stack1&0724810101 072481301\\
		&FullStack&0112520601 0112520701 0112520901 0200470101 0561580401 0724810101 0724810301\\
		
		Holmberg IX X-1 &&0200980101\\
		&FullStack&0112521001 0112521101 0200980101 0693850801 0693850901 \\
		&& 0693851001 0693851101 0693851701 0693851801\\
		
		IC 342 X-1$^{1}$&Stack1&0693850601 0693851301 \\
		
		M33 X-8&FullStack&0102640101 0102641801 0141980501 0141980801\\		
		
		NGC 1313 X-1&Stack1&0405090101 0693850501 0693851201\\
		&Stack2&0803990101 0803990201 0803990501 0803990601\\
		&Stack3&0803990301 0803990401 0803990701\\
		
		NGC 1313 X-2&Stack1&0150280301 0150280401 0150280501 0150280601 0150281101 0205230301 0205230601\\	
		&Stack2&0764770101 0764770401\\	
		&FullStack&Stack1+Stack2\\
		
		NGC 247 ULX&Stack1& 0844860101 0844860201 0844860301 0844860401 0844860501 0844860601 0844860801\\
		&Stack2&same as Stack1 but different broadband continuum\\
		
		NGC 300 ULX-1&&0791010101\\
		&&0791010301\\
		
		NGC 4190 ULX-1&FullStack&0654650101 0654650201 0654650301\\
		
		NGC 4559 X-7&&0842340201\\
		
		NGC 5204 X-1&FullStack&0142770101 0142770301 0150650301 0405690101 0405690201 \\
		&& 0405690501 0693850701 06938501401 0741960101\\
		
		NGC 5408 X-1$^1$&Stack1&0653380201 0653380301\\		
		&Stack2&0653380401 0653380501\\
		&FullStack&0302900101 0500750101 0653380201 0653380301 0653380401 0653380501\\
		
		NGC 55 ULX&&0655050101\\
		&&0824570101\\
		&&0864810101\\
		&FullStack&0655050101 0824570101 0864810101\\
		
		NGC 5643 X-1&&0744050101\\
		
		NGC 6946 X-1&&0691570101\\
		
		NGC 7793 P13 &Stack1&0804670201 0804670301 0804670401 0804670501 0804670601 0804670701\\
		&FullStack&0693760401 0781800101 0804670201 0804670301 0804670401\\
		&& 0804670501 0804670601 0804670701 0823410301 0840990101\\
		
		RX J0209.6-7427&&0854590501\\
		
		SMC X-3 &&0793182901\\

		\hline
	\end{tabular}\\
	$^{1}$ Another source in the source extraction region, the data could be partly contaminated.
	
\end{table*}

\section{The cross-correlation method}
\label{method}

It is not computationally expensive to perform a systematic automated search for Gaussian lines in an X-ray spectrum if one just wants to locate the strongest residuals in the spectrum and find their \delcstat\ fit improvement. The search however gets much more expensive if it is necessary to establish the true significance of these features  including the look-elsewhere effect. Given that each automated Gaussian search can take of the order of one hour to perform, the need to perform thousands of searches on simulated datasets easily results in the requirement of 10000 computer hours to run the search on a single object. This is not an unreasonable time to expend in a study of a single object, but the method quickly becomes prohibitively expensive if we want to study a larger source sample. We therefore needed to improve the search method.

To decrease the required computational time, we employ a cross-correlation approach. For two discrete arrays, their cross-correlation $C$ takes a simple form of:

\begin{equation}
\label{crosscorr}
C=\sum^{N}_{i=1} x_{i}y_{i}
\end{equation}

\noindent where $x$ and $y$ are two arrays of real numbers of the same length $N$. In principle the cross-correlation can also be applied to arrays of unequal lengths but using the same lengths simplifies the problem. From equation \eqref{crosscorr} we can see that if the two arrays have similar values at the same array elements, their cross-correlation will be large. If their values at the same array elements are dissimilar (e.g. random noise centred on zero), the cross-correlation will be small. If the values are similar but of a different sign, the cross-correlation will have a large absolute value but it will be negative.

Therefore, if we are searching for Gaussian lines in a spectrum, we could imagine fitting it with a broadband continuum spectral model, printing the flux residuals to this fit into an array and then cross-correlating these residuals with an array containing the spectral model of a Gaussian line with predefined parameters such as the line position (wavelength) and the line width. Then the parameters of the Gaussian line could be changed and the new model could be again cross-correlated with the ULX spectral residuals. The Gaussian parameters could be changed in an automated fashion following a grid of line positions (=wavelengths) and line widths (equivalent to the velocity width of plasma due to turbulent or rotational motion). We would therefore obtain the cross-correlation value of the dataset residuals to a moving Gaussian of any (reasonable) parameters.

\citet[][Sections 2.1, 2.2]{Zucker+03} finds that under some conditions, the likelihood is an increasing monotonic function of the squared cross-correlation. Therefore a maximum (or a negative minimum) of the cross-correlation function will maximize the likelihood - i.e. the Gaussian of specific parameters which maximizes the cross-correlation value will also maximize the likelihood of the fit. In other words, these are the best-fitting parameters of such Gaussian to the dataset residuals. The conditions required are that both arrays need to be continuum subtracted: 

\begin{equation}
\sum^{N} x_{i}=\sum^{N} y_{i}=0
\end{equation}

\noindent This is approximately satisfied by the residual dataset since the X-ray spectrum is fitted by the best-fitting continuum model. In case of strong emission or absorption complexes in the source spectrum, the best-fitting continuum will lie between the true broadband continuum and the residuals, so that the fitting statistic, $\chi^2$ or C-stat, is minimized (thus roughly satisfying the condition above). The spectral model (Gaussian) array can easily be shifted by a constant amount such that the condition above is satisfied.

Therefore we can use the value of the cross-correlation function versus the Gaussian parameters to find the best-fitting position of a Gaussian if fitted to the residual source spectrum. However, an important problem appears here. The value of the strongest cross-correlation (at the best-fitting Gaussian parameters) will not tell us directly how much the Gaussian line fit is preferred to the null hypothesis and what is the probability that any residual originated purely from noise. Furthermore, even if it did, it would still not include the look-elsewhere effect - the fact that we searched a broad space of parameters (line widths and wavelengths) to find the preferred solution.

These issues can be solved if we perform the same cross-correlation search on the residuals of fake spectra simulated from the best-fitting source continuum spectrum but containing just Poisson noise. This is the same approach as used in the direct fit search methods described in Section \ref{intro_oldmethods}. By performing the same search on simulated data, we obtain a distribution of cross-correlation values for each tested Gaussian parameter. Therefore we can say how unusual is the cross-correlation value seen in the real dataset for such Gaussian parameters, and by extension what is the false positive rate of this cross-correlation value. This gives us the significance of any line detection if we performed just a single trial. In the following text we name this quantity the single trial significance.

To take into account the look-elsewhere effect, we have to `equalize' the searches at different Gaussian parameters - the cross-correlation value could mean something completely different at one wavelength in comparison with another wavelength. As can be seen from Eq. \ref{crosscorr}, the cross-correlation value takes into account only the absolute values of the residual flux and the Gaussian flux, and ignores the uncertainties on the flux. This means that a residual at a certain wavelength in the dataset will produce a stronger cross-correlation value than another emission residual with a lower absolute flux regardless of the size of uncertainties on the individual flux data points. Therefore the first residual might mistakenly appear more significant than the second one even if the uncertainties on its flux data points are much larger than those on the second residual. The required `equalization' of different searches must be achieved by a re-normalization of the cross-correlation values so that the values are equivalent for different Gaussian parameters.

A cross-correlation search with a Gaussian of specific wavelength and line width on simulated datasets will produce a distribution of cross-correlation values centred approximately on $C=0$. In general, Gaussians with different wavelengths ($\lambda$) and different line widths ($w$) will produce different cross-correlation distributions, however they all originate from the same Poissonian noise process so their shapes should be equivalent. We can therefore rescale the cross-correlation distributions at different Gaussian parameters to be equivalent, using the statistics from the simulated datasets. 

The choice of the renormalization formula is not obvious. We choose the following renormalization factor $R_{\lambda,w}$ for each Gaussian parameter $\lambda,w$:

\begin{equation}
R_{\lambda,w}=\sqrt{\frac{1}{N}\sum^{N}C_{i}^{2}}
\end{equation}

\noindent where the sum is over all the simulated datasets with the same Gaussian parameters $\lambda$ and $w$. Therefore $R_{\lambda,w}$ is equivalent to the standard deviation $\sigma$ of distribution $C$ if its mean is equal to zero (which should be approximately the case). We thus define the renormalized cross-correlation value such that:

\begin{equation}
RC=\frac{C}{R_{\lambda,w}}
\end{equation}

\noindent where $\lambda$ and $w$ are the parameters of the Gaussian with which the cross-correlation was obtained and $C$ is the raw correlation value. This quantity then indicates how unusual each cross-correlation value is in units of $\sigma$ in the simulated datasets, regardless of its Gaussian parameters. This renormalization also removes the dependence of our results on the Gaussian line normalization (both the raw cross-correlations and $R_{\lambda,w}$ scale linearly with line normalization) - the line normalization can thus be fixed to any value in our search.

Now, if our choice of renormalization factor was correct, this quantity should be equivalent to the single trial significance obtained for its Gaussian parameters $\lambda, w$. However, importantly, the maximum of the $RC$ value is not limited by the number of simulated searches we performed as opposed to the single trial significance.

We take one further step here. The Poisson distribution generating the noise in our problem is not completely symmetric around the zero value of the residual, i.e. on the negative side the residuals can only reach down to zero X-ray flux but on the positive side there is no limit to how strong a residual can be. The exact shape of the positive and negative cross-correlation distributions can thus be slightly different. This difference likely decreases with increasing data quality, as the Poisson distribution becomes more symmetric, approaching the Gaussian distribution. We therefore split the simulated cross-correlation distributions for each Gaussian parameter into positive distributions (raw cross-correlation larger than 0) and negative distributions (raw cross-correlation lower than zero) and calculate their renormalization factors independently. The renormalized correlation of a positive residual is then

\begin{equation}
RC_{+}=\frac{C_{+}}{R_{\lambda,w+}}=\frac{C_{+}}{\sqrt{ \frac{1}{N_+} \sum C_{i+}^{2}}}
\end{equation}

\noindent where the sum in the denominator is only over all the positive raw cross-correlations in the simulated distribution (at Gaussian parameters $\lambda$, $w$). The renormalized correlation of a negative residual is calculated in the same manner but only summing all the squares of the negative raw cross-correlations in the simulated distribution.

The renormalization puts the searches with all the different Gaussian parameters on equal footing. This means we can now compare them. Now, finally, to calculate the true false positive rate (and significance) of any line detection in the real dataset, we need to compare its $RC$ value with all the simulated $RC$ values at any Gaussian parameters. The false positive rate is the fraction of simulated spectral residuals which produce at least one $RC$ value (at any Gaussian parameter) larger than the one seen in the real dataset.

The steps of the cross-correlation analysis are as follows: 

\begin{enumerate}
\item The real source spectrum is reduced from the raw dataset. 
\item The spectrum is fitted with a broadband spectral model, and the residuals to this model are recovered. \item Any low quality wavelength bins are identified in the data and ignored. 
\item We generate the simulated datasets based on the broadband model to the continuum and their residuals to the model, 
\item We generate the (Gaussian) spectral models for any parameter (wavelength or line width/velocity width) of interest. \item The residuals and the spectral models are cross-correlated (each dataset with each spectral model) and we obtain the raw cross-correlations. 
\item The raw cross-correlations are renormalized and we recover the $RC$ values (for both real and simulated data) and the resulting true significances for each Gaussian parameter in the real dataset. 
\item Finally, we select only the most significant line features in the source spectrum for the final ULX line catalogue. The only criterion for selection was the true significance of any cross-correlation peak. We selected all lines with true significance above 1$\sigma$. This cutoff corresponds to a lower limit of line single trial significance of around 3$\sigma$ (exact value varies for different sources).
\end{enumerate}

All of the individual steps are explained in more detail in Appendix \ref{methodappendix}. The analysis gives us three different quantities to assess the significance of any detection:
\begin{itemize}
\item The single trial significance ($STS$) defines how unusual is the cross-correlation value seen in the data compared with simulated searches of the same Gaussian properties. $STS$ naturally ignores the look-elsewhere effect.
\item The renormalized cross-correlation ($RC$) should be approximately equivalent to the single trial significance but is not limited by the number of simulations as it is calculated from the distribution of raw cross-correlation values rather than from their order. It also does not take into account the look-elsewhere effect. 
\item The true significance ($TS$) is calculated from the true false positive rate and indicates the true probability that a feature seen in the real dataset originates from Poisson noise, including the look-elsewhere effect. The true false positive rate is determined by comparing the $RC$ values at all searched Gaussian parameters. $TS$ will underestimate the detection significance for spectral lines which are not Doppler-shifted because it assumes the worst case scenario (a line with any shift, any reasonable width, anywhere in the observed spectrum).
\end{itemize}

\section{Results}
\label{results}

\subsection{The performance of the cross-correlation method}

The computational performance of the method was tested on a desktop computer powered by a quad-core Intel processor. As the method requires frequent loading and saving of files, it strongly benefits from using local storage. At the same time, using large blocks of simulated datasets (e.g. 5000 per file) allows for non-local storage as well, at the cost of reduced performance and increased RAM memory requirement (16 GB required).

We find that the whole automated cross-correlation search on a single source takes 1 to 2 hours to run on the test computer if one performs 10000 dataset simulations and searches roughly 2000 wavelength bins (accurately sampling the RGS spectral resolution) and 12 different Gaussian velocity widths (ranging from 250 km/s to 5000 km/s). The time required depends on the number of wavelength bins in the search (i.e. how finely we search for Gaussian lines), the number of velocity width bins, the spectral binning of all the spectra searched in the real dataset, as well as on how many simulations are performed in the search. We chose to perform 10000 simulations per source to balance the computational cost and reasonably high maximum achievable significances (a false positive rate of 1 in 10000 corresponds to a significance of $3.9\sigma$).

In comparison, the traditional Gaussian line search where the line is fitted directly within \textsc{spex} takes of the order of one hour to scan a single RGS spectrum (real or simulated). We have thus achieved a speed-up of the Gaussian spectral search by roughly a factor of 10000 to 100000.

\subsection{The accuracy of the cross-correlation method}

We also tested the accuracy of the new method. We found a clear correlation between the normalized cross-correlation and the single trial significance in all the datasets searched. On average, the relative difference between these two quantities, that is the standard uncertainty of the ratio $\frac{STS}{RC}$, was between 1 and 4 \%, and decreased with increasing data quality. Such a small difference suggests that the choice of the normalization factor $R_{\lambda,w}$ was reasonable and that the renormalized correlation is a very good indicator of how unusual is each residual in its own wavelength bin. 

However, the range of renormalized correlations is not limited by the number of performed simulations as opposed to the single trial significance. Renormalized correlation is calculated from the sum of the squared raw correlations within each parameter bin rather than by counting the simulated correlations stronger than the real data (within each bin). In other words, $RC$ takes into account the shape and the size of the raw cross-correlation distribution in each parameter bin rather than just the fraction of simulated cross-correlations in the extreme wing (beyond the raw cross-correlation value of the real dataset). The renormalized correlation can therefore indicate a higher significance than the single trial significance at the same number of performed simulations, which is why we prefer it.

Comparing the cross-correlation method with the direct fitting method, we find a clear correspondence between the normalized correlation and the $\Delta$C-stat fit improvement obtained from directly fitting the strongest lines in \textsc{spex}. However, the scatter between these two quantities is larger than in the case of normalized correlation versus single trial significance. The scatter can likely be attributed to the fact that the two methods (direct fitting and cross-correlation) are based on completely different principles.

To make a valid comparison between the direct fitting and the cross-correlation method, we compared them on a controlled sample. We simulated ULX RGS spectra and searched them with both methods using the same Gaussian parameter grids. We simulated and searched three types of source spectra: 50 RGS spectra, each with $\sim10^6$ source counts, representing a very high quality high-resolution dataset; 50 RGS spectra with $\sim10^4$ source counts representing a good quality ULX dataset (based on the FullStack NGC 5204 X-1 spectrum) and 50 RGS spectra with $\sim10^3$ source counts, representing a lower quality ULX dataset (based on the Stack2 approach on NGC 1313 X-2). The Gaussian parameter search grid had a wavelength spacing of 0.01 \AA\ (same as in the real ULX search) and we used a single velocity width bin of 1000 km/s.

Each step of the direct fitting search procedure involves a fit of a spectral model composed of the original continuum plus a Gaussian with a fixed width and wavelength. It is therefore a spectral fit of one extra free parameter compared with the original continuum fit. The statistics improvement compared with the original fit can be denoted \delcstat. Then the statistical significance of the added spectral component (Gaussian line) with such parameters is roughly equal to $\sqrt{\Delta \textrm{C-stat}}$. As shown by \citet{Cash+79}, the C-stat difference between these two spectral models has approximately the form of the $\chi^2$ function (with one degree of freedom). The $\sqrt{\Delta \textrm{C-stat}}$ quantity is therefore the single trial significance of adding the extra Gaussian line and thus is roughly equivalent to the renormalized cross-correlation value in the cross-correlation method.

\begin{figure*}
	\centering
	\includegraphics[width=\textwidth]{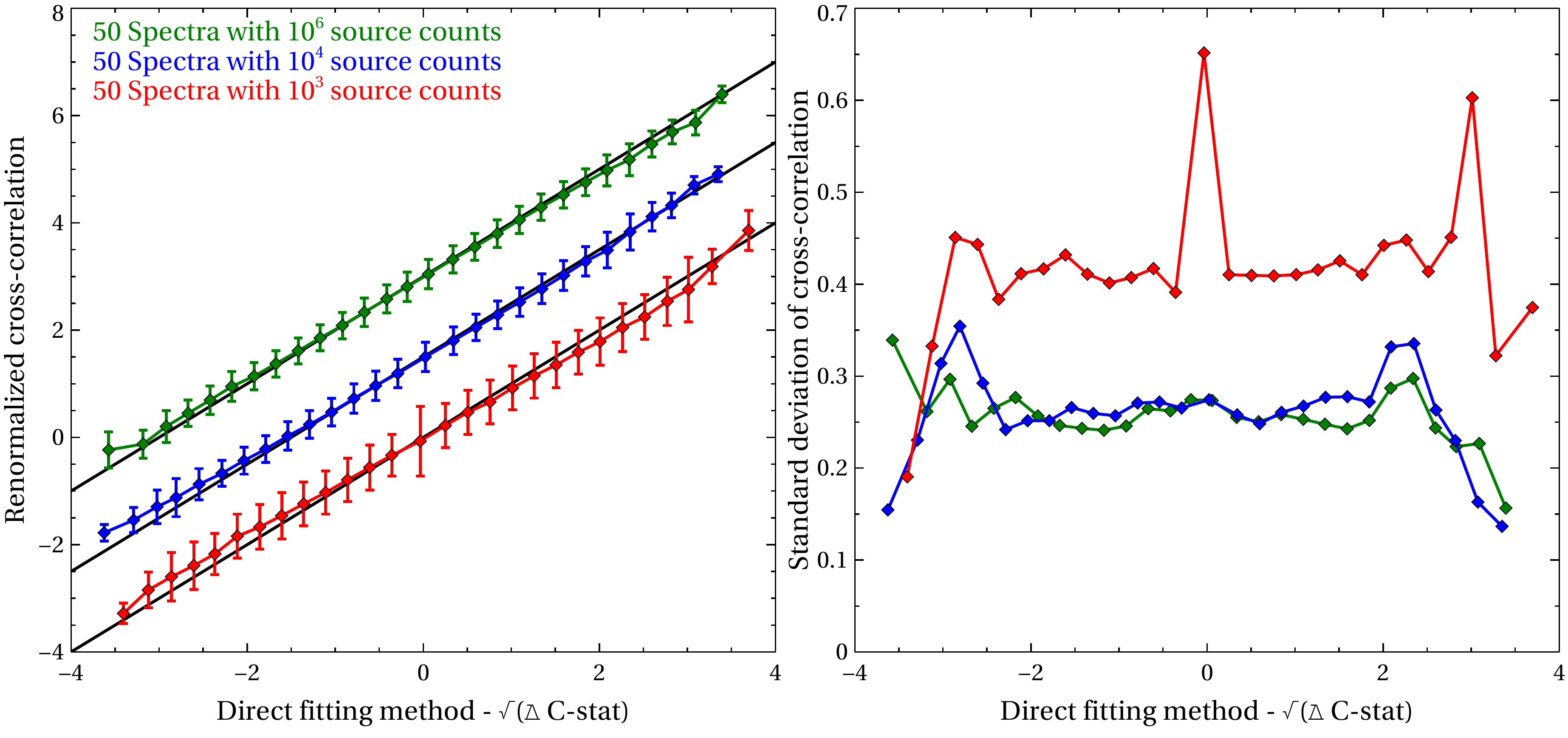}
    \caption{Comparison between the direct fitting and the cross-correlation methods. 150 spectra of 3 different data qualities (shown here in different colours according to the legend) were simulated and searched with both methods. The left subplot shows the comparison of two equivalent quantities from both methods for all searched parameters from all the simulations. The blue and green groups were offset vertically for a better visualisation, the black lines correspond to y=x functions for the corresponding data groups. The original clouds of points were adaptively binned horizontally so that each point has Gaussian statistics (minimum 25 data per point) and so that they sample the horizontal range by roughly 0.25. The uncertainty of each bin is the standard deviation of cross-correlation values within it. The right subplot shows the standard deviation of the cross-correlation distribution across the horizontal (direct fitting method) range.}
    \label{crosscorr_traditional_comparison}
\end{figure*}

Fig. \ref{crosscorr_traditional_comparison} (left) shows the comparison between $\sqrt{\Delta \textrm{C-stat}}$ and the renormalized cross-correlation at the same Gaussian parameters in the simulations of various dataset quality. We find very good agreement between the average results from the two search methods, throughout the full range of $\sqrt{\Delta \textrm{C-stat}}$ explored and for all three different dataset qualities. At the same time, we find that there is random scatter between the results from the different methods. The absolute standard deviation between the methods appears stable for any $\sqrt{\Delta \textrm{C-stat}}$ (Fig. \ref{crosscorr_traditional_comparison} - right) and is $0.2-0.3$ for the higher quality datasets, while being $0.4-0.5$ for the lower quality $10^3$ source count dataset. Thus the relative deviation between the methods decreases for stronger residuals, reaching approximately 10\% relative errors at the significance of $\sim3\sigma$. We particularly note that all the spectral residuals of relevance will be located at these extreme ends of the distribution where the relative difference is lowest. We consider this an acceptable difference between the two methods considering that they are based on completely different principles and conclude that the new method is reasonably accurate for use in this study.

We also checked the cross-correlation search results from the 150 synthetic simulations for the presence of false detections due to any possible RGS instrumental features (e.g. many detections at the same wavelengths in the simulated spectra). We did not find any such features.

\subsection{Example analysis: NGC 1313 X-1}

\begin{figure*}
	\centering
	\includegraphics[width=0.92\textwidth]{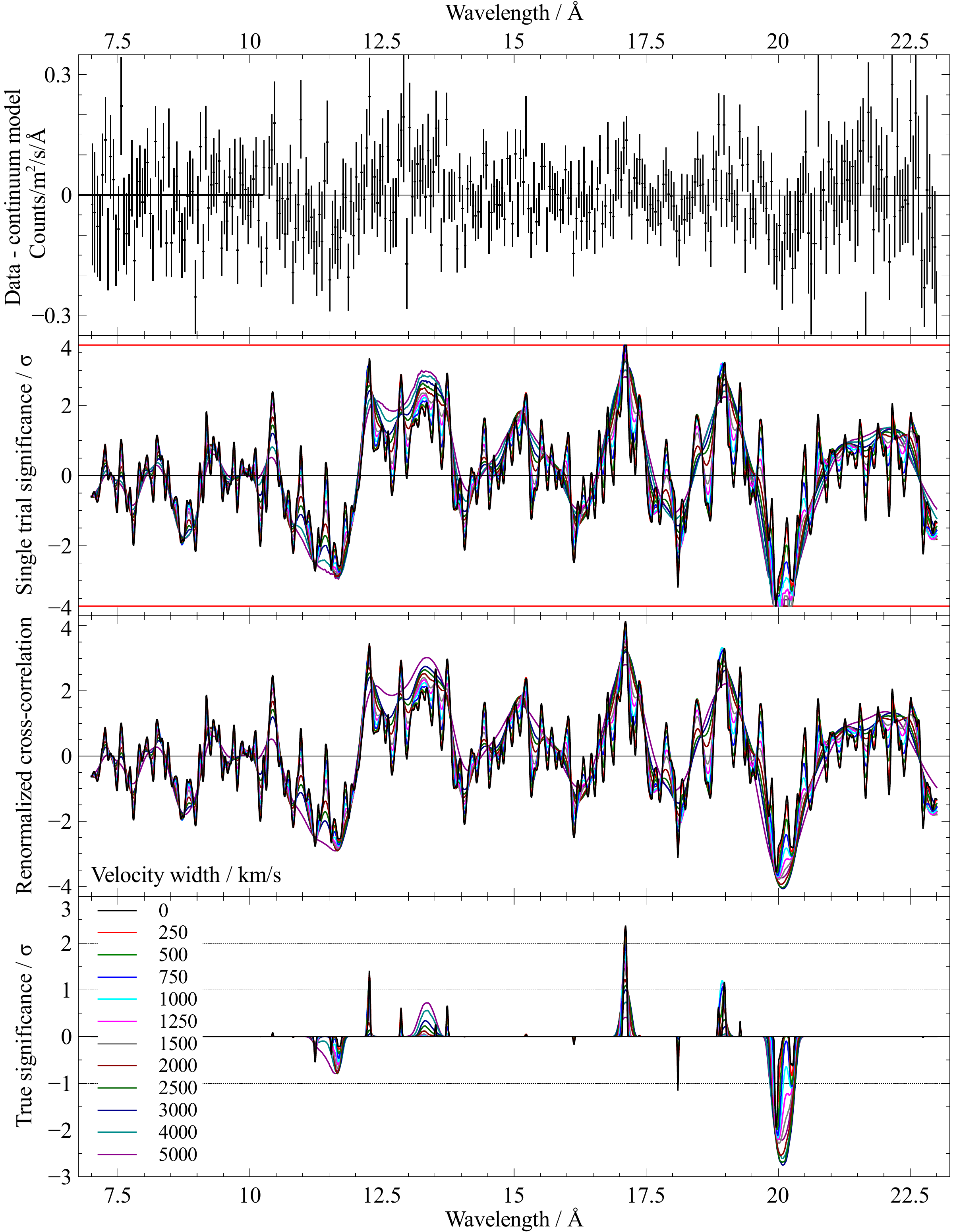}
    \caption{The cross-correlation search of Stack1 of NGC 1313 X-1 performed between 7 and 23 \AA. The top subplot shows the (stacked and heavily over-binned) RGS residuals to the broadband spectral continuum. The second subplot contains the single trial significance, the renormalized correlation is shown in the third subplot and the true significance is in the bottom subplot. Searches with different Gaussian velocity widths are in different colours (according to the plot key in the bottom subplot). The red horizontal lines in the second subplot show the minimum/maximum attainable single trial significance given the performed number of simulations (10000).}
    \label{NGC1313spectrum}
\end{figure*}

\begin{table*}
	\centering
	\caption{Excerpt from the ULX catalogue containing just the strongest lines detected in the Stack1 observations of NGC 1313 X-1. The 20-column catalogue table is split into 4 rows for display purposes. Each column is described in more detail in Appendix \ref{cataloguestructure}.} 
	\vspace{0.3cm}
	\label{NGC1313table}
	\begin{tabular}{ccccc}
		\hline
		Object name&Approach&Wavelength&Energy&Turb. velocity\\
		&&\AA\ &keV&km/s\\
		\hline
		
		NGC1313X1& Stack1& 1.2260e+01& 1.0113e+00&0.0000e+00\\
		NGC1313X1& Stack1& 1.7110e+01& 7.2463e-01 &0.0000e+00\\ 
		NGC1313X1& Stack1& 1.8100e+01& 6.8500e-01& 0.0000e+00\\
		NGC1313X1& Stack1& 1.8940e+01& 6.5462e-01& 1.0000e+03\\
		NGC1313X1& Stack1& 2.0090e+01& 6.1715e-01& 3.0000e+03\\ 
		\hline
		True p-value &   True signif. &   Renorm. corr.& Single trial p-value & Single trial sig.\\
		\hline
	 	1.5910e-01& 1.4081e+00& 3.4639e+00& 8.1284e-04& 3.3484e+00\\
	 	  1.7500e-02& 2.3760e+00& 4.1371e+00& 2.0092e-04 &3.7179e+00\\ 
	 	  2.4740e-01& -1.1567e+00& -3.1172e+00& 1.4028e-03& -3.1941e+00\\
	 	  2.2890e-01& 1.2032e+00& 3.3313e+00& 1.6191e-03& 3.1524e+00\\
	 	  6.0000e-03& -2.7478e+00& -4.0637e+00& 1.9759e-04& -3.7221e+00\\
		\hline
		\delcstat\  &  Photon flux &  -    &         + & En. flux \\
		&ph/cm$^2$/s & ph/cm$^2$/s &  ph/cm$^2$/s & erg/cm$^2$/s   \\
		\hline
		1.0990e+01& 3.0283e-06 &-9.7796e-07 &1.0407e-06& 4.9049e-15\\
		 1.4320e+01 &2.1361e-06& -6.4111e-07& 6.6604e-07 &2.4803e-15\\
		 1.1030e+01& -2.0115e-06& -5.5441e-07& 5.7847e-07& -2.2071e-15\\
		 1.0540e+01& 2.7839e-06& -8.8956e-07& 9.7828e-07& 2.9199e-15\\ 
		 1.4800e+01& -6.0691e-06 &-1.5135e-06 &1.5523e-06& -6.0031e-15\\
		\hline
		-        &     +       &     Equiv. width   & -         &    +\\
		erg/cm$^2$/s   &        erg/cm$^2$/s      & keV & keV & keV \\
		\hline
	 	-1.5840e-15& 1.6856e-15& 5.9537e-03& -2.0084e-03& 2.1334e-03\\
	 	-7.4443e-16& 7.7338e-16& 3.6281e-03& -1.1412e-03&1.1845e-03\\
	 	-6.0831e-16& 6.3472e-16& -2.9891e-03& -8.6689e-04& 9.0348e-04\\
	 	-9.3301e-16& 1.0261e-15& 4.2088e-03& -1.4055e-03& 1.5408e-03\\
	 	-1.4970e-15& 1.5354e-15 &-1.0183e-02& -2.6860e-03& 2.7540e-03\\
		\hline
	\end{tabular}\\
	
\end{table*}

We show an example analysis of the archetypal ULX NGC 1313 X-1 with the cross-correlation method as this ULX exhibits a known ultra-fast wind \citep{Pinto+16, Pinto+20}.

We present the analysis of Stack1, which is the original dataset where \citet{Pinto+16} discovered the ionised outflow in absorption as well as ionised rest-frame emission. The EPIC pn and RGS datasets, reduced following Section \ref{datareduction} are fitted with the standard three component ULX broadband spectral continuum described in Section \ref{continuummodelling}. We find a powerlaw slope of 2.18, the temperature of the cooler blackbody is 0.17 keV and the temperature of the hotter blackbody is 3.39 keV. All three components are obscured by neutral absorption with a column of $2.18\times10^{21}$ \pcm.

We generate the Gaussian spectral models with the different velocity widths (according to Section \ref{modelgeneration}) in the useful wavelength range of 7 \AA\ to 23 \AA, which is not background-dominated. Then the source spectra are simulated using the continuum model and the same clean exposure of 287 ks as the real dataset. Afterwards the models and the residuals are cross-correlated following Section \ref{methodcrosscorr}. We derive the renormalized correlation, true significance and the single trial significance for each wavelength bin of the searched range, and for each velocity width in our parameter grid. These quantities are shown in Fig. \ref{NGC1313spectrum} alongside the raw RGS residuals. A comparison can be made with Fig. 3 in \citet{Pinto+16}. The differences between the results can be attributed to differences in the search methods used and in the chosen broadband spectral continuum.

We immediately notice a strong absorption residual at 20 \AA\ (interpreted as O VII absorption blueshifted by $\sim$0.1c). The single trial significance plot shows that this residual is so strong that none of the 10000 simulations produced a comparably strong residual at that specific wavelength, even though its true significance is just below 3 sigma. This shows how important it is to account for the look-elsewhere effect for features not found at the rest-frame wavelengths of any expected transitions. There is also a broad absorption residual at $11-12$ \AA, which appears weak in the true significance plot, however a broader absorption feature (or multiple lines) would likely fit the residual better (and with a much higher significance). \citet{Pinto+16} show that this residual particularly stands out when a broader velocity width is used (10000 km/s). This is beyond the scope of this project, which focuses mainly on narrow line features.

Additionally, we also notice a number of emission features, especially at wavelengths of 12, 17 and 19 \AA. These correspond to rest-frame emission from the ions of Ne X, Fe XVII and O VIII. Their minimum significances are between $1-3\sigma$ however since they likely correspond to rest-frame emission, their real significance is more in line with the value of the renormalized cross-correlation or the single trial significance.

We caution the reader against direct comparisons of this Gaussian line scan with more in-depth searches using physical plasma models which prove that the ionised outflow detection in NGC 1313 X-1 is highly significant at $4-5\sigma$ \citep{Pinto+20}. The plasma models aggregate significance by combining the fit improvement statistics of multiple spectral lines at once, which all agree with the same outflow scenario. Therefore while a single feature might appear insignificant by itself, a combination of lines at different wavelengths, all fitted with the same physical model can result in a strong detection of an outflow (which is the case here). 

The information about the individual spectral features is condensed in the ULX line catalogue where we only list the significantly detected lines ($>1\sigma$ true significance). Table \ref{NGC1313table} reports an excerpt from the catalogue showing just the search of this dataset, containing the strongest features.

\subsection{The full sample}

The final catalogue of the strongest detected features (with true significance above 1$\sigma$) contains 135 spectral lines, of which 82 are emission and 53 are absorption lines.

We have obtained the true p-value for each line, i.e. the maximum false positive rate in case the feature is not located at a wavelength of any expected atomic transition (which includes the look-elsewhere effect). This means that we can directly calculate the maximum contamination rate of the catalogue - the probability that an average feature from the catalogue originates due to noise rather than due to a physical process. This percentage, obtained by summing all the individual line true p-values and dividing by the size of the catalogue is found to be 11\%. We find that the contamination fraction of emission features, 10\% (at most $\sim8$ fake emission features in the catalogue), is somewhat smaller than that of absorption features, which is 13\% (at most $\sim7$ fake absorption features in the catalogue). Assuming instead that all the emission features originate from rest-frame plasma (which might not be the case) reduces the contamination fraction of emission lines down to only 0.06\% ($<<1$ expected fake emission feature in the catalogue). This once again illustrates how important is the look-elsewhere effect in a blind search of a high-resolution dataset.

\begin{figure*}
	\includegraphics[width=\textwidth]{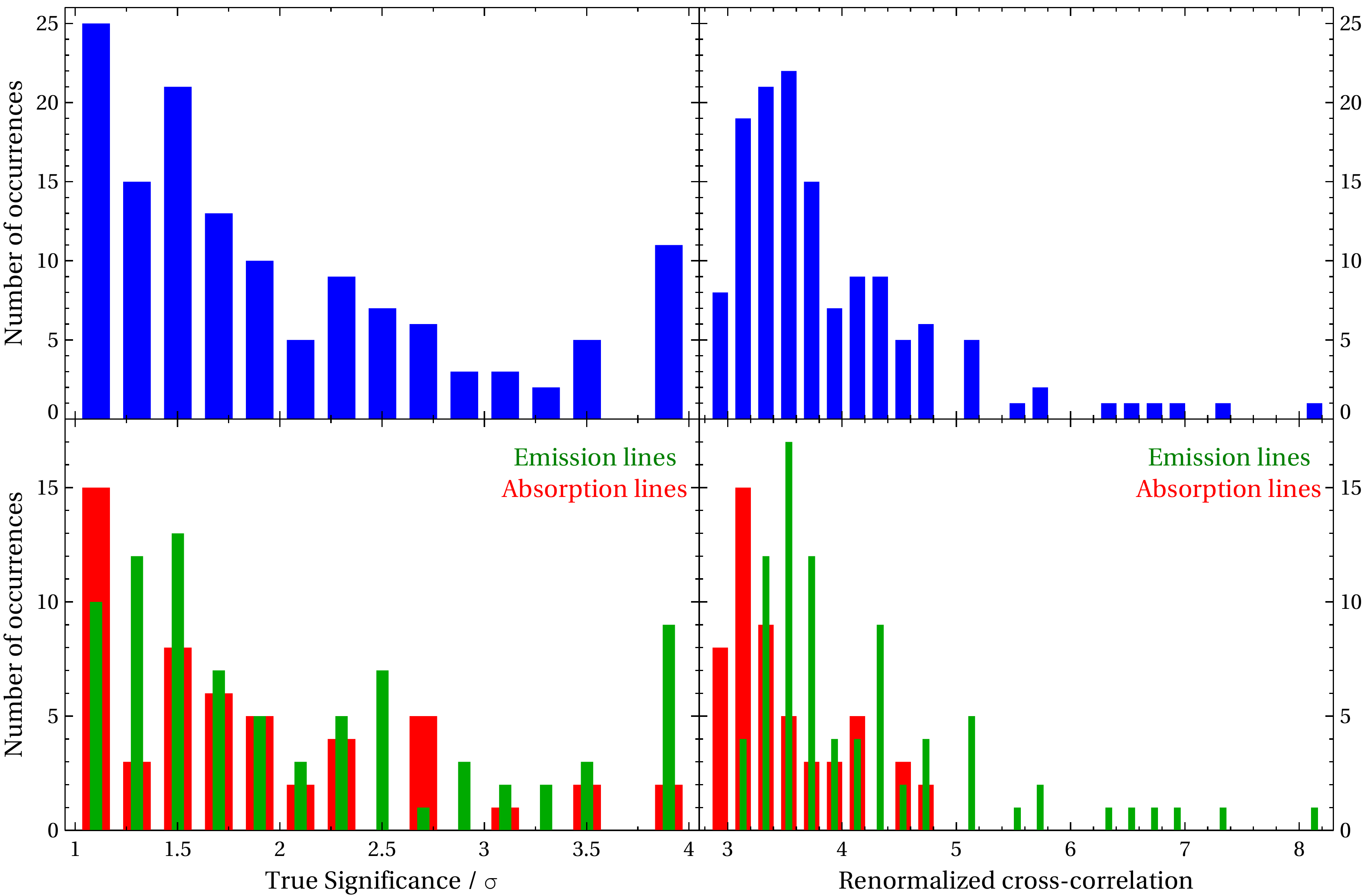}
    \caption{The histograms of true significances (left subplots) and renormalized cross-correlations (right subplots) of the detected lines. The top subplots show the total statistics for all lines, while the bottom subplots show the statistics for emission and absorption lines separately.}
    \label{sigcorr_hist}
\end{figure*}

To study the statistics of the detected lines, we created histograms of their significances. The true significance and the renormalized cross-correlation distributions are shown in Fig. \ref{sigcorr_hist}. The lower cutoff at true significance of 1$\sigma$ is imposed by our selection criteria. The peak of true significances near 4$\sigma$ is due to the total number of simulations performed per object giving the maximum achievable significance (i.e. in a number of these cases, the $\sim4\sigma$ significance quoted is actually a lower limit to the actual detection significance). We notice that many of the detected features apparently have quite low true significances between 1 and 2$\sigma$. However, it is important to note that these are the absolute minimum significances of these features in the case that they are not located near an expected elemental transition (which many are expected to be). Even though the significances might seem low, the overall catalogue contamination fraction is not large at about 11\%. If we study the emission and absorption feature statistics separately (lower subplots in Fig. \ref{sigcorr_hist}), we find their distributions are reasonably similar except for a higher abundance of lower significance absorption features, and a lack of very significant ($\sim4\sigma$) absorption features.

\section{Discussion}
\label{discussion}

\begin{figure*}
	\includegraphics[width=\textwidth]{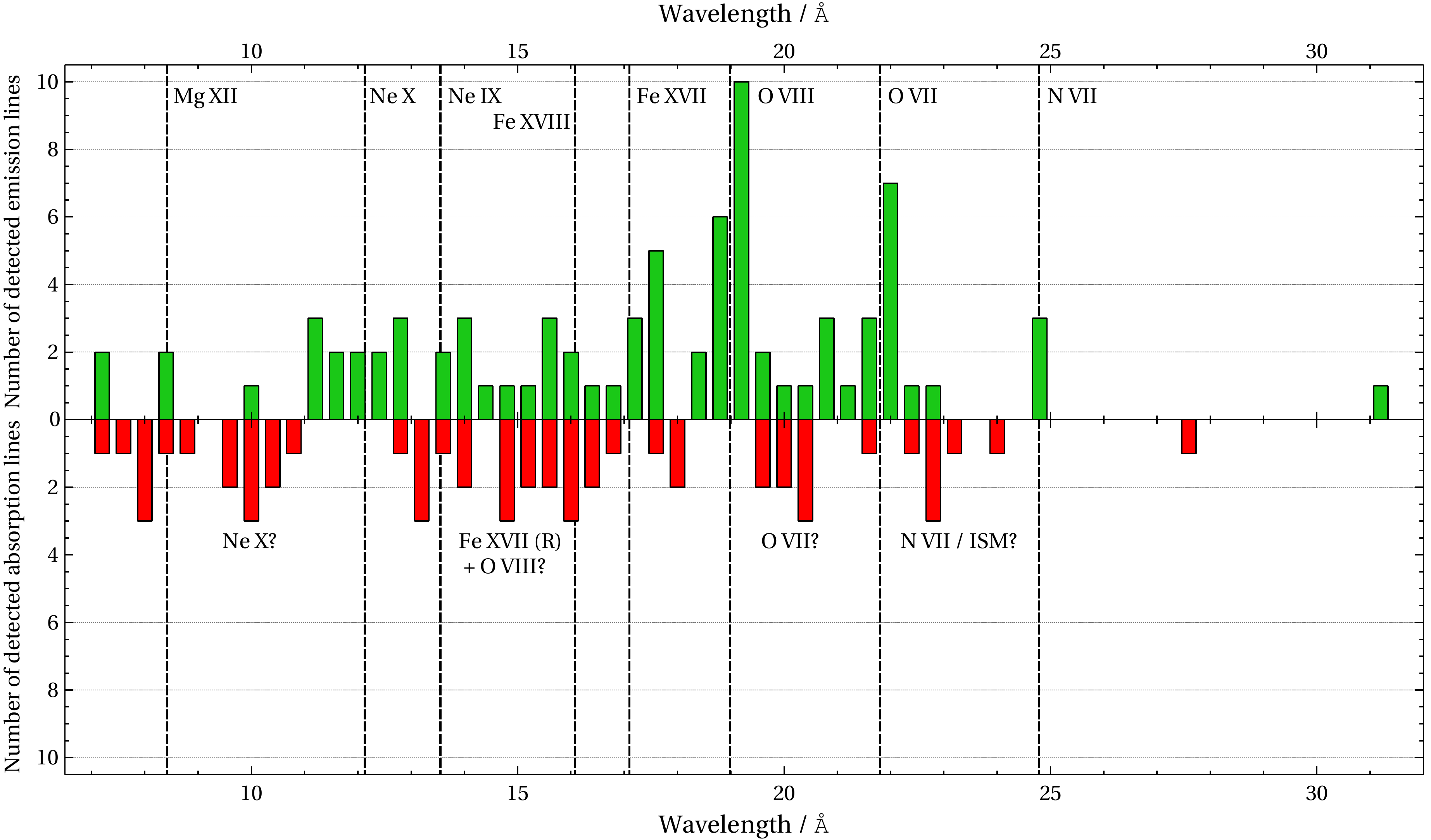}
    \caption{The histograms of the emission (green) and absorption (red) lines detected in the full sample versus their wavelength. The histograms are binned by 0.4\AA. Labels show the likely identification of the most abundant emission lines and the vertical dashed lines give the rest-frame wavelengths of these transitions. Considering the absorption lines are most likely Doppler-shifted, their preliminary identifications must be taken cautiously.}
    \label{wave_hist}
\end{figure*}

In this work, we collected all suitable high resolution \xmm\ RGS data of ULXs and of two nearby super-Eddington pulsars and searched them for ionised plasma spectral features, both in absorption or emission. Collecting the 135 strongest line detections (with rigorously determined detection significances), we created the first catalogue of spectral lines in ULXs. Up to this point nothing was assumed about the origin and the emission/absorption process which produced these spectral features. In attempt to understand their physics, we plot the wavelengths of the significantly detected emission and absorption lines separately in two histograms, shown in Fig. \ref{wave_hist}.

As expected, we find that many of the detected emission lines are grouped around known strong elemental transitions. This has been previously remarked by \citet{Pinto+16} and \citet{Kosec+18a} but using much smaller samples of ULXs and line detections. The most commonly observed features are the emission lines of O VII (rest-frame wavelength of the triplet around 22 \AA) and O VIII (19 \AA). There is also strong evidence for Fe XVII/XVIII, the wavelengths of the strongest lines of its species are around 15-16 \AA\ and then particularly around 17 \AA. Another strongly detected element is Ne, represented by Ne IX (around 13.5 \AA) and Ne X (12.1 \AA). The range around $11-12$ \AA\ could also possibly contain the emission lines from rest-frame Fe XX-XXIV transitions. Finally, we also observe detections around the N VII transition (24.8 \AA) and some evidence for the Mg XII transition at 8.4 \AA.

The situation is completely different if we consider the absorption features. In general, we find that not many absorption lines occur at wavelengths with common occurrence of emission features (the rest-frame positions of strong elemental transitions). This is particularly true for the very common O VII, O VIII and Ne X features, although the Fe XVII/Fe XVIII region between 13 and 17 \AA\ seems to be an exception with presence of both emission and absorption lines. The observed anti-correlation of occurrence of absorption and emission features is not surprising, considering that the absorption likely originates from a fast disc wind crossing our line of sight towards the central X-ray source. These winds have been shown to flow at large velocities ($0.1-0.3$c) in a few ULX \citep{Pinto+16, Pinto+17, Kosec+18b}, hence the blueshifts of the absorption lines are considerable. If these winds indeed occur in most ULXs and with such typical velocities, we can tentatively guess the identification of the detected absorption residuals.

The absorption lines appear to be clustered into multiple groups. The group observed around 20 \AA\ could originate from blueshifted O VII absorption (with velocities of $\sim$0.1c). The broad group seen between 14.5 \AA\ and 17 \AA\ could then be a blend of O VIII and Fe XVII-XVIII absorption with velocities of $0.1-0.2$c. The next strong group is at $9-11$ \AA\ and could originate from Doppler-shifted Ne X absorption (again shifted by $\sim$0.2c, with possible contribution from slower Fe XX-XXIV absorption). The groups around $8$ \AA\ and $13-14$ \AA\ could be imprinted by fast Fe XXIV and Fe XVII/XVIII ions, or by Mg XI/XII and Ne IX if the projected wind velocities are somewhat lower. Finally, we also observe a group of features between 22 and 24 \AA. These could originate from blueshifted N VII absorption (at $\lesssim0.1$c), however this wavelength range also contains a number of low ionisation O lines (e.g. O II and III at 23.4 and 23.0 \AA, respectively) and dust absorption features \citep[$22.8-23.0$ \AA, e.g.][]{Pinto+13} which could be imprinted on the ULX spectrum by the intervening interstellar medium (the continuum spectral model only accounts for the neutral gas).

\begin{figure*}
	\includegraphics[width=0.8\textwidth]{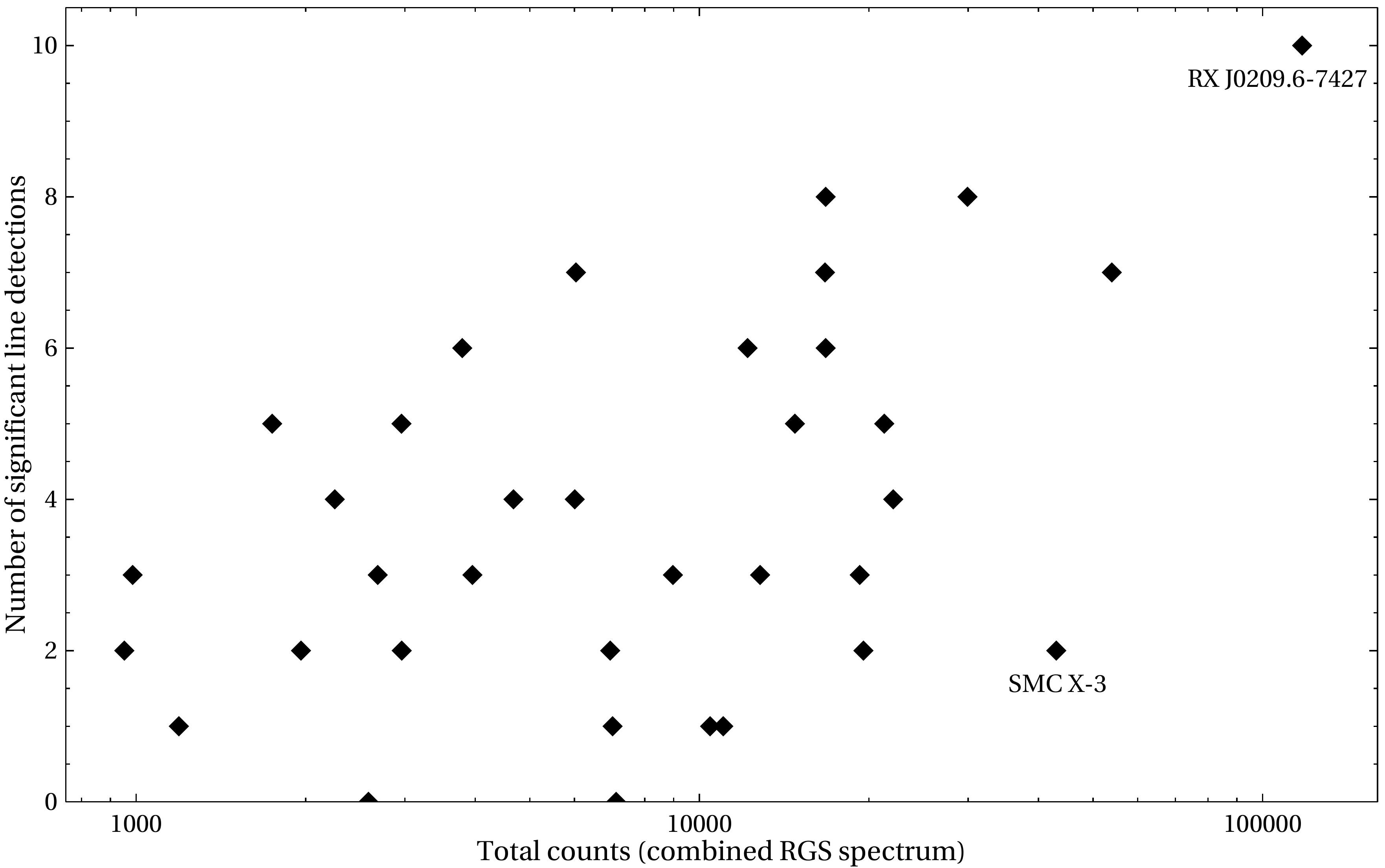}
    \caption{The number of significantly detected lines in each individual approach versus the number of net counts in its combined RGS spectrum. Labels show the super-Eddington pulsars in our sample. The remaining points all correspond to ULXs.}
    \label{detections_vs_counts}
\end{figure*}

We also compare the results of spectral searches of different sources. It is particularly interesting to compare the number of significantly detected lines versus the ULX data quality (RGS counts) and other properties such as its spectral hardness or its X-ray luminosity. ULXs show a large range of spectral hardnesses \citep{Sutton+13}, thought to be related to their inclination angles and/or their mass accretion rates. The number of significantly detected features versus the quality of source spectra is shown in Fig. \ref{detections_vs_counts}. Naturally, we find that better data quality on average results in more significant detections.

\begin{figure*}
	\includegraphics[width=\textwidth]{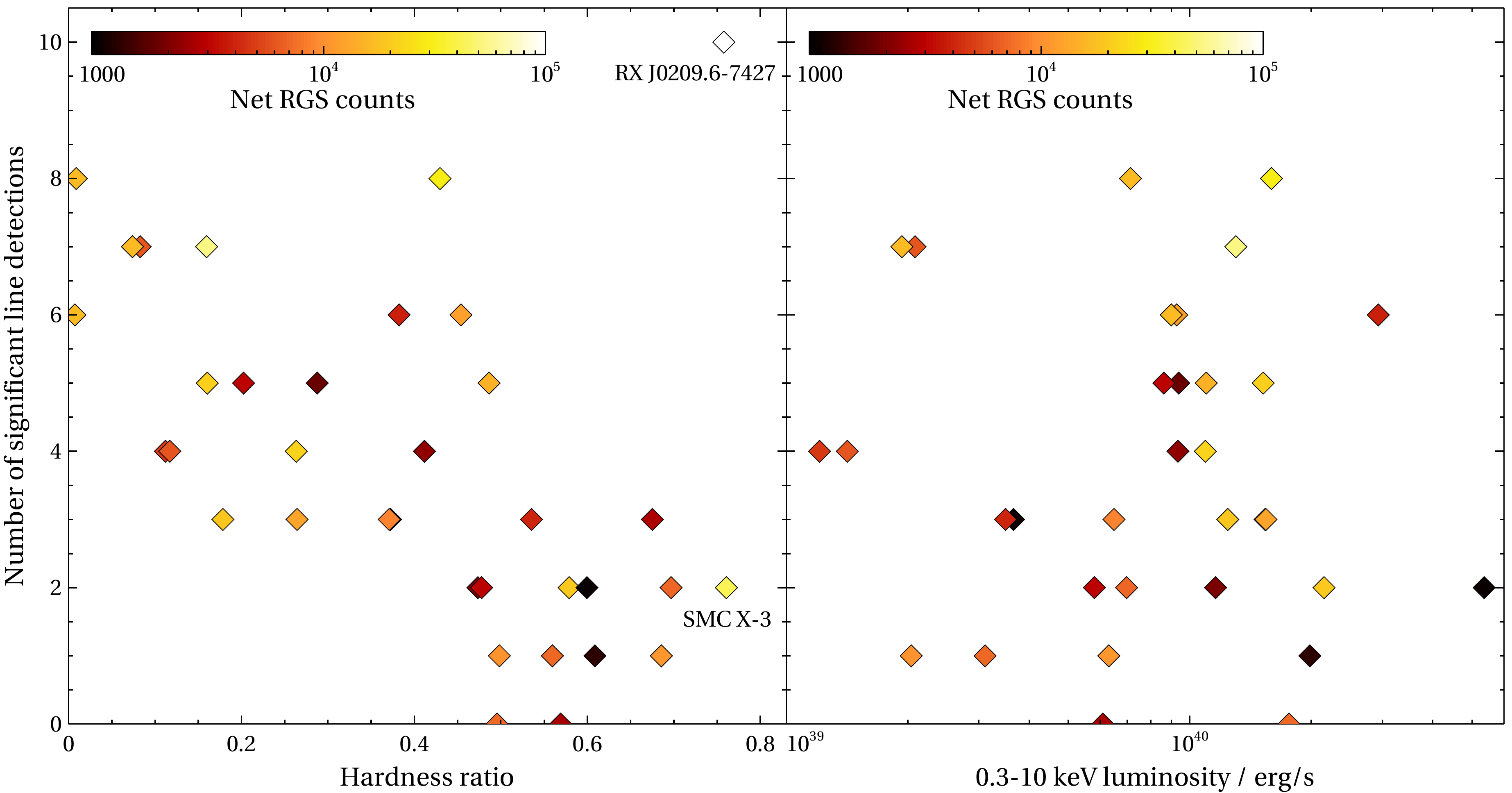}
    \caption{Left subplot: The number of significantly detected lines in each individual approach versus the spectral hardness of the source calculated from the broadband spectral model such as H/(H+S), where H is the $2-10$ keV de-absorbed luminosity and S the $0.3-2.0$ keV de-absorbed luminosity. Right subplot: The number of significantly detected lines in each individual approach versus the de-absorbed $0.3-10$ keV luminosity calculated from the broadband spectral model. The colour scale in both subplots shows the total net counts in the combined RGS spectrum.}
    \label{detections_vs_hardness_vs_totlum}
\end{figure*}

Importantly, we also find that spectrally harder ULXs show fewer detections than spectrally soft ULXs (Fig. \ref{detections_vs_hardness_vs_totlum} - left subplot). The colour scheme in Fig. \ref{detections_vs_hardness_vs_totlum} shows the data quality (number of source counts in the combined RGS spectrum), and illustrates that even good quality RGS datasets ($\sim$10000 counts) of hard ULXs result in few line detections while much lower quality soft ULX datasets often show many more line detections. Similar results were previously presented in \citet{Kosec+18a} but using a much smaller sample of ULXs and a different analysis method. The Pearson correlation coefficient of the relationship between the number of line detections and the spectral hardness (the two super-Eddington pulsars excluded) is $-0.67$ with a false positive probability of $2.4\times10^{-5}$, suggesting a highly significant anti-correlation. To show that this is not a data quality effect, we split the ULX-only sample by data quality into two groups. The higher data quality group gives a Pearson coefficient of $-0.62$ (p-value $9.1\times10^{-3}$) and the lower data quality group a coefficient of $-0.74$ (p-value of $6.6\times10^{-4}$). 

We also studied whether this anti-correlation holds for emission and absorption lines separately. Naturally the statistics of the separate populations are much smaller and thus the trends are weaker. Importantly, we found that none of the two line populations show an equally strong trend as seen in the combined dataset. This suggests that a similar anti-correlation is observed in both emission and absorption line populations. The Pearson correlation coefficients for these two populations (with the two super-Eddington pulsars excluded) are $-0.56$ (p-value $8.7\times10^{-4}$) for emission lines and $-0.59$ (p-value $4.3\times10^{-4}$) for absorption lines.

One of the leading scenarios explaining the difference between spectrally soft and hard ULXs suggests that soft ULXs are very similar objects to hard ULXs but observed from higher inclination angles. In that case the hotter (and spectrally harder) central accretion flow regions are obscured from our view in soft ULXs by a geometrically thick super-Eddington accretion disc but directly visible in the hard ULXs. A schematic of this scenario is shown in Fig. 13 of \citet{Pinto+17}. The scenario can readily explain the lack of absorption features in hard ULXs since the ionised disc wind might not be crossing our line of sight in these sources at all. In fact \citet{Pinto+20a} find a correlation between the projected velocity, the ionisation parameter of the outflow and the hardness ratio of the ULX for the few sources in which fast winds were detected. This correlation was interpreted as an orientation effect.

However, to explain the lack of emission lines is more challenging. If the hardness is directly related to object inclination (and no other quantities), the emission line regions should be subject to the same spectral energy distribution (SED) in both hard and soft sources since their position with respect to the accretion flow should not change. Thus they should produce the same radiation output as they originate from optically thin plasma. At the same total luminosity the harder sources have less flux in the soft X-ray band where the lines are detected than the soft sources. The contrast between the lines and the continuum should then be even higher and they should be easier to detect in harder ULXs. Alternatively, the emission lines could be outshined in harder ULXs by the directly visible inner accretion flow regions (contributing also to soft flux), leading to lower line equivalent widths (and harder detectability) as suggested by \citet{Middleton+15} and \citet{Pinto+17}. This however necessarily implies a higher X-ray luminosity but we find no correlation between the number of line detections and the ULX luminosity (Fig. \ref{detections_vs_hardness_vs_totlum} - right). Hence the lack of emission lines is difficult to reconcile unless they are obscured in hard ULXs, which seems unlikely. Perhaps it suggests that other factors such as the mass accretion rate are more important drivers of ULX spectral hardness rather than their orientation towards us alone.

Therefore the observed anti-correlation between the number of features and the source hardness suggests some difference between the plasma conditions or location in soft and hard ULXs. 
The difference in line detection rates could be due to the different spectral energy distributions (SEDs) of these two subclasses of ULXs. The harder SEDs of hard ULXs could ionise the plasma elements to higher ionisation levels than the softer SEDs of soft ULXs, resulting in weaker line features which are much harder to detect (particularly in ULXs with poorer RGS data quality). Alternatively, if radiation line pressure contributes or drives the outflows, harder SEDs would result in less driving force and thus in lower mass outflow rates (although even soft ULX SEDs are quite hard for radiation line driving). Future ULX studies with instruments such as \textit{Athena} achieving many more line detections (in much less exposure time), especially above 1 keV (from hotter plasma phases), will likely be able to explain this difference between soft and hard ULXs.

Interestingly, no correlation is seen between the number of significantly detected features and the observed source X-ray ($0.3-10$ keV) luminosity, derived from its continuum spectrum. This is shown in Fig. \ref{detections_vs_hardness_vs_totlum} (right subplot). For emission lines, the lack of correlation indicates that the ratio of emission line luminosity to the $0.3-10$ keV source luminosity does not change dramatically (considering there is no correlation between the object luminosity and the RGS data quality), and hence the mass of the X-ray-emitting plasma scales with the luminosity of the object. This is observed despite the mass of the accreting system not scaling with the X-ray luminosity (assuming these objects are all stellar-mass accretors). For absorption lines, the lack of any correlation with luminosity suggests little evolution of the absorber optical depth with luminosity. The fact that we observe a similar set of absorption lines indicates a similar ionisation parameter of plasma, and thus the absorber column density must remain roughly constant, unless the absorber only partially covers the source and the lines are saturated \citep[as seen in quasars,][]{Hamann+19}. However, the ionisation parameter $\xi$ is related to the source ionising luminosity $L_{ion}$ such that 

\begin{equation}
\xi=\frac{L_{\rm{ion}}}{nR^2}
\end{equation}

\noindent where $n$ is the plasma density and $R$ the absorber distance from the ionising source. Hence either $n$ or $R$ must increase to compensate for the increased luminosity. Increased density at a constant column density leads to a thinner absorption layer and also a constant mass outflow rate. This seems unlikely as the wind mass outflow rate likely scales with the mass accretion rate (and hence the luminosity). It therefore appears that the absorption distance $R$ must be increasing with increasing luminosity. We note that \citet{Pinto+20} also observe an increase in the wind launching radius in NGC 1313 X-1 with increase in its luminosity. Thus it appears that both of these dimensions rise with ULX luminosities despite no scaling in the physical sizes of their accreting systems. Alternatively, if the lines are saturated, the absorption distance does not need to increase, but the partial covering factor cannot significantly evolve with ULX luminosity.

There may be alternative explanations for the observed anti-correlation between the number of detected lines and ULX spectral hardness, and the lack of correlation with luminosity, beyond different inclinations and mass accretion rates. Two black hole ULXs, despite similar luminosities, could have different fractions of mass lost to outflows and energy lost to advection/photon trapping, leading to very different line spectra and hardnesses (due to down-scattering in the wind). Similarly, two neutron star ULXs with comparable luminosities could have different spins and magnetic field strengths, resulting in different magnetosphere sizes. Assuming the outflow is produced in a supercritical part of the disc beyond the magnetosphere, different magnetosphere size would lead to very different outflow properties. Smaller magnetosphere would likely result in more massive and faster outflows due to the supercritical disc extending more inwards. The different magnetosphere size would also lead to different spectral hardnesses considering the accretion disc is spectrally softer than the accretion column and more mass in outflows would likely result in more photon down-scattering.

So far we mostly studied the sample of ULXs and super-Eddington pulsars as whole. From Fig. \ref{detections_vs_hardness_vs_totlum} (left), where the two pulsars are specifically labelled, we can see that they are the hardest sources in our sample. SMC X-3 only shows two significant line detections, in line with the trend seen in spectrally harder ULXs. On the contrary, RX J0209.6-7427 shows 10 line detections despite its hardness (the most per source in the whole sample), however we note that its RGS dataset is by far the highest quality dataset in the sample with over 100000 combined source counts (see Fig. \ref{detections_vs_counts}). The statistics of detected line wavelengths are very limited, but we observe strong similarities with the full sample. Most emission lines are seen near strong rest-frame transitions (O VIII, O VII and N VII, however the higher ionisation lines such as Fe XVII and Ne X are missing), while most absorption lines are avoiding the expected transition wavelengths.

A single spectral feature (especially if it has a $1-2\sigma$ minimum significance) detected alone in a ULX spectrum is not equivalent to an ionised outflow detection. However, if a single feature is detected, it is likely that any possible plasma/outflow might have also imprinted other spectral features, potentially weaker and thus not selected by our procedure. In Fig. \ref{detections_vs_counts} we can see that there are four cases of a single significant line detection in a dataset.

The natural next step is therefore to try to describe multiple spectral lines at once, using a physical plasma model (ionised emission or absorption). The plasma model can be generated for a broad range of plausible physical parameters such as the ionisation parameter, systematic velocity and velocity width, and the ULX spectra can be searched for its spectral signatures. The significance of any plasma detection will be a combination of the significances of the individual spectral lines. Even features weak individually can add up to a significant detection because the true significance of a detection rises very steeply with increasing fit improvement \delcstat\ (or $\Delta \chi^2$). For example, in the case of NGC 300 ULX-1 \citep{Kosec+18b} where a fast ionised outflow was detected with a true significance of around $3.7\sigma$, its strongest single spectral line (blueshifted O VIII line) was found in this work to be only significant at 1.3$\sigma$ alone (true significance). Applying a physical model can thus reveal much weaker plasma signatures. On the other hand, an important disadvantage of searching the ULX spectra with these models is that this is necessarily a much more model-dependent approach than using simple Gaussian line models to describe the residuals. Alternatively, a simpler and less model-dependent compromise could be to adopt a combination of a small number of Gaussian lines - e.g. an emission line triplet (to describe O VII or Ne IX emission), or to use a P-Cygni shape (to describe a possible wide angle outflow contributing both to ionised emission and absorption).

It is possible to extend the current cross-correlation analysis to physical plasma models or more complex spectral shapes by replacing the spectral model generation step in Section \ref{modelgeneration}. This extension is beyond the scope of this paper and will be addressed in future work.

\section{Conclusions}
\label{conclusions}

We systematically studied all suitable high-resolution soft X-ray spectra of Ultraluminous X-ray sources and searched them for plasma signatures in emission or in absorption. To assign the true false positive probability to each of the detected features (including the look-elsewhere effect), we developed a new, computationally affordable method of searching for Gaussian features in X-ray spectra. The method is based on cross-correlation and it is more than 10000 times faster than the previous approaches based on automated direct spectral fitting. By collecting all the detected spectral features, we created the first catalogue of spectral line detections in the soft X-ray spectra of ULXs. The catalogue contains 135 candidate lines (82 emission, 53 absorption lines) with a contamination fraction due to noise of at most 11\%. Over 90\% of studied sources show at least 1 spectral line, and roughly a third of the sources at least 5 line detections.

Most detected emission features are located at wavelengths corresponding to known transitions of ionic species of O, Fe, Ne, N, and Mg. On the other hand, the absorption lines generally appear to avoid these wavelengths and instead appear to be distributed between them. This is in agreement with a hypothesis that the emission lines originate in low-velocity material, while the absorption lines originate in fast disc winds crossing our line of sight with velocities of 0.1 to 0.3c. If this is indeed the case, such ultra-fast outflows are common in many ULXs alongside with lower-velocity wind components producing the observed emission lines.

We also find that spectrally harder ULXs show fewer spectral line detections than spectrally soft ULXs. This indicates a difference in the ULX accretion geometry/viewing angle which cannot be explained purely by their different orientation towards the observer. The difference could be due to over-ionisation of the ionic species by the harder spectral energy distributions of harder ULXs. Further observations with high-resolution X-ray instruments are necessary to increase the line statistics and understand the observed trend. At the same time, no correlation is observed between the number of line detections and the ULX X-ray luminosity.

Further research directions for the systematic approach pioneered in this work include the extension of the cross-correlation method. It could be extended to allow the automated search for physical plasma models in X-ray spectra. Another option is the application of the search method to other X-ray sources such as active galactic nuclei and Galactic X-ray binaries. The method could in particular be applied for automated search of ultra-fast outflows in the X-ray spectra of active galactic nuclei.

\section*{Acknowledgements}

We are grateful to the anonymous referee for useful comments which improved the quality of the manuscript. P.K. acknowledges support from the European Space Agency. Support for this work was provided by NASA through the Smithsonian Astrophysical Observatory (SAO) contract SV3-73016 to MIT for Support of the Chandra X-Ray Center (CXC) and Science Instruments. C.S.R. thanks the UK Science and Technology Facilities Council for support under the New Applicant grant ST/R000867/1, and the European Research Council for support under the European Union's Horizon 2020 research and innovation programme (grant 834203). This work is based on observations obtained with XMM-Newton, an ESA science mission funded by ESA Member States and USA (NASA). This research has made use of the NASA/IPAC Extragalactic Database (NED), which is funded by the National Aeronautics and Space Administration and operated by the California Institute of Technology.


\section*{Data Availability}

All of the data underlying this article are publicly available from ESA's \xmm\ Science Archive\footnote{https://www.cosmos.esa.int/web/xmm-newton/xsa} and NASA's HEASARC archive\footnote{https://heasarc.gsfc.nasa.gov/}.




\bibliographystyle{mnras}
\bibliography{References} 




%

\appendix

\section{Catalogue structure}
\label{cataloguestructure}

The catalogue is in the form of a single table saved in the ASCII format. The table contains all the strongest emission and absorption features as selected from the raw results, which have a true significance of at least 1$\sigma$. The table contains 135 rows and 20 columns plus a 2-row header with column descriptions and physical units. Each row corresponds to a single line feature detected in a specific ULX. The columns contain the following. The first column contains the source name and the second the observation or observations in which the line feature was found. Column 3 lists the wavelength of the feature (in \AA), column 4 the energy (in keV) and column 5 the velocity width (in km/s) at which the line shows the strongest cross-correlation. Columns 6 and 7 contain the true false positive rate and significance of the feature. Column 8 lists the renormalized correlation of the feature, and columns 9 and 10 show its single trial false positive rate and the single trial significance. Column 11 lists the \delcstat\ fit improvement value obtained upon directly fitting the feature in the \textsc{spex} fitting package. Columns 12 to 14 contain the line photon flux (in photons/cm$^2$/s) and its lower and upper errorbars. Columns 15 to 17 contain the line energy flux (in erg/cm$^2$/s) with lower and upper errorbars, respectively. Finally, columns 18 to 20 contain the line equivalent width (in keV) and its lower and upper errorbars.

We note that wherever the true significance, the renormalized cross-correlation and the single trial significance are negative in the catalogue, this simply indicates that the feature found is an absorption line. Positive values of those quantities indicate that the feature is an emission line. We also note that all of the uncertainties in the catalogue are stated at 1$\sigma$ confidence.

\section{Step-by-step explanation of the cross-correlation method}
\label{methodappendix}

Analysis parts \ref{datareduction} through \ref{spectralmodelgeneration} as well as part \ref{collection} are run mostly as \textsc{bash} scripts (because \textsc{bash} offers automated access to \textsc{spex}) but also involve manual fitting within \textsc{spex} (part \ref{continuummodelling}). Part \ref{methodcrosscorr} is written in \textsc{python}.

\subsection{Data reduction}
\label{datareduction}

In the first step, all \xmm\ data are reduced. The data were downloaded from the \xmm\ Science Archive and reduced using the standard SAS v17.0.0 pipeline, CALDB as of June 2020. We reduced \xmm\ RGS and EPIC pn data.

The RGS data were reduced using the \textsc{rgsproc} procedure, and filtered for any flaring events with a threshold of 0.25 cts/s in each of the detectors. They were binned by a factor of three directly within the \textsc{spex} fitting package to oversample the instrumental resolution by roughly a factor of three. They were used in the wavelength range which was not dominated by background flux. The exact range depended on the source and background fluxes of each object but often the range between 7 \AA\ and 20 or 26 \AA\ was chosen. For approaches using multiple stacked observations, we stacked the RGS 1 and RGS 2 spectra separately, producing two independent spectra to be fitted simultaneously.

The EPIC pn data were used to model the broadband (0.3-10 keV) ULX spectrum correctly. The data were reduced using the \textsc{epproc} procedure and filtered for any background flares with a threshold of 0.5 cts/s. The images of the pn exposures were prepared with the \textsc{evselect} procedure. The source regions were circles, usually with a radius of 35 arcsec. However, this was not always possible due to contamination by nearby X-ray sources. In those cases a smaller source region radius (20 or 25 arcsec) was chosen. The background regions were of polygon shape, at least 110 arcsec away from the main source to avoid the wings of its point spread function, at the same time avoiding any other bright X-ray sources. The background regions were located on the same chip as the source and were as large as possible to maximize the background statistics, whilst still located in the Copper hole on the EPIC pn chip \citep{Freyberg+04, Carter+07}. The background-subtracted pn spectra were binned to at least 25 counts per bin (achieving Gaussian statistics) and to oversample the real spectral resolution by a factor of at most three using the \textsc{specgroup} procedure. They were used in the 0.3 to 10 keV spectral range, but ignored in the wavelength range where RGS data were available. This way the spectral fit was not driven by EPIC pn data with much higher statistics (but much poorer spectral resolution) in the useful RGS range.

\subsection{Continuum modelling}
\label{continuummodelling}

After data reduction, the source spectra (RGS1 + RGS2 + pn) were fitted with a broadband continuum spectral model, to locate any potential residuals around the best-fitting X-ray continuum. The fitting was performed manually within the \textsc{spex} fitting package to avoid mis-fitting of the spectra.

We chose to use a phenomenological ULX spectral model, previously employed also by \citet{Kosec+18a}. The model is composed of three emission components: a powerlaw, a blackbody modified by coherent Compton scattering and a standard blackbody. The powerlaw component (\textsc{pow} in \textsc{spex}) represents emission from the innermost regions of the ULX accretion flow - from an optically thin corona in the case of a black hole accretor or from an accretion column in the case of a neutron star accretor. The second component, a blackbody modified by coherent Compton scattering (\textsc{mbb} in \textsc{spex}), with temperatures of 1 to 3 keV represents X-ray emission from the hot, inner accretion disc of the ULX. The third component, a standard blackbody (\textsc{bb} in \textsc{spex}) with lower ($0.1-0.2$ keV) temperatures represents either emission from the colder, outer accretion disc or emission from an optically thick outflow launched by the super-Eddington accretion flow. Finally, the spectrum is affected by interstellar absorption along our line of sight towards the ULX. Neutral gas in both our Galaxy and in the ULX host galaxy can contribute to this absorption, and thus the absorber column density was left free to vary in our continuum fits.

The spectral model is motivated by the physical picture outlined above. Nevertheless, the best-fitting model parameters such as the blackbody temperatures and powerlaw slopes should be interpreted with caution as our models do not include description of the hard X-ray ULX emission (above 10 keV). The absence of hard X-ray data (by only including \xmm\ spectra) could lead to systematic uncertainties on these continuum parameters. However, ultimately, the model is primarily designed to ensure a good phenomenological fit to the \xmm\ continuum, and therefore should not be compared too seriously to the results from truly broadband spectral fits which include the \nustar\ hard X-ray coverage.

The broadband model fits the spectra of most ULXs very well with no obvious broad systematic residuals. Considering the ULX spectral hardness is calculated from the model luminosity (rather than X-ray flux), a model-dependent hardness ratio uncertainty could be introduced, if the model over- or under-predicts the true neutral absorption column density. This would result in systematically over- or under-predicted ULX hardness ratios (with no effect on the actual Gaussian search). However, we do not estimate the errors introduced through the model choice to be too serious and prefer this approach to calculating spectral hardness from raw fluxes, which ignore the various ULX column densities altogether.

In a minority of cases, the soft standard blackbody was not required for an acceptable fit. In those cases only the powerlaw and the modified blackbody was used. This was particularly the case in those ULXs where the neutral absorption column was high (above $10^{22}$ \pcm).

\subsection{Pre-filtering the data}
\label{prefilter}

In the third step of the routine we identified any low quality wavelength bins in the source spectrum and discarded them. In the current version of the cross-correlation method we simply discarded any wavelength bins where the continuum spectral model value was abnormally high - usually these defect bins appear as a delta function in the spectral plot and are too narrow to be real lines in the continuum spectral model. They correspond to bad pixels on the RGS detectors. The filtering threshold was chosen manually upon inspection of the spectrum, and the identified wavelength bin positions were discarded automatically in further analysis. The testing of our method showed that not excluding these defects could affect the correct renormalization of the cross-correlations later in the analysis.

\subsection{Generating real and simulated residual spectra}
\label{modelgeneration}

After the low quality bins were identified, the residual spectrum of the source around the best-fitting broadband model was generated and saved. The Y-axis of the spectrum was in units of Photons/m$^2$/s/\AA, i.e. the residuals were saved in physical units. The exact Y-axis unit is unimportant but it must be a unit of flux rather than a ratio to model values or a ratio to error bars, otherwise the cross-correlation method would find peaks in the data quality space rather than fitting physical shapes.

The source spectrum was simulated with the \textsc{simulate} command within \textsc{spex}, assuming just the best-fitting broadband continuum affected by Poisson noise, and assuming the exposure of the same length as the real spectrum (and with the same background level). The residuals of each simulation to the continuum model were saved similarly as done with the real dataset. We neglect the uncertainties on the assumed broadband continuum, however these are unlikely to be significant considering the continuum is anchored by the high quality EPIC spectrum and is smooth and featureless.

The simulations were repeated as many times as required. In this study we performed 10000 simulations for each source, which means that we can probe significances of up to about 4$\sigma$. We found that the best computational performance was achieved if the simulated residuals were stored in files by large blocks, for example by storing 5000 individual simulations in a single file. This grouping results in a large table where the columns are individual simulations and the rows correspond to the same wavelength bins. As we are searching through data from two individual instruments (RGS 1 and RGS 2), each column begins with data from RGS1, followed by data from RGS2.

\subsection{Generating spectral models}
\label{spectralmodelgeneration}

The next step was to generate the spectral models to cross-correlate with the real and simulated datasets. First the real dataset was loaded into \textsc{spex} (RGS1 and RGS2 simultaneously) and the low quality wavelength bins were ignored, thus ensuring the wavelength bins and range were identical to the ones in the real and simulated data. Then the spectral model to be searched for was loaded. In this work, we simply loaded a Gaussian line as the spectral model. The Gaussian had a predefined line width (calculated from the velocity width $v$ we were searching for) and a predefined wavelength $\lambda$. The velocity width of the line is related to its full width at half maximum through the following equation: $FWHM=2.355v\lambda/c$. The value of its normalization was positive (it was an emission line) but its exact value was unimportant (considering the cross-correlations were later renormalized) and was kept constant for all spectral searches in this study. The spectral model was saved, and then the parameters of the Gaussian were varied according to a grid of parameters we were searching over, saving the model at each step. Each saved spectral model was a column with as many rows as there were wavelength bins in our RGS 1 and RGS 2 datasets. The spectral models for RGS1 and RGS2 were similar but not exactly the same (they have different defect pixels and chips, different chip gap wavelengths and slightly different effective areas).

We searched the full usable RGS wavelength range (the exact range depended on each specific source) with a precision of 0.01 \AA, which slightly oversampled the instrumental resolution of RGS. We also probed a range of different velocity widths of plasma producing the spectral lines. The line width was calculated from the appropriate velocity width at each wavelength. As a compromise between sampling and computational expense, we searched 12 different velocity widths: 0, 250, 500, 750, 1000, 1250, 1500, 2000, 2500, 3000, 4000 and 5000 km/s. We did not search for lines with widths too large to avoid interpreting broad continuum model residuals, likely originating in imperfect broadband modelling, as absorption or emission lines. For each source dataset, the spectral models were stored in 12 individual table files, one for each value of velocity width.

\subsection{Cross-correlation}
\label{methodcrosscorr}

After the real/simulated data and the spectral model files were obtained, the main cross-correlation part of the routine was performed. First, all the spectral model file tables were loaded into RAM memory as a 3D array.

Secondly, the real source dataset was cross-correlated with the all spectral model files, file-by-file, and column-by-column in each model file. We used the \textsc{correlate} function within the \textsc{numpy} package in \textsc{python} programming language. Cross-correlation is a symmetrical process and therefore even though our model files were composed of emission Gaussian lines, we were searching simultaneously for emission (positive correlations) and absorption (negative correlations) features.

Afterwards, the simulated datasets were loaded block-by-block and cross-correlated with all the spectral model files in the same fashion as done with the real data, and their raw correlations were saved. We calculated the number of each positive and negative correlation simulations (in each parameter bin), as well as the sums of squares of all positive and negative correlations (independently) in each bin.

When all the simulated dataset blocks were processed, the positive and negative normalization factors for each bin were calculated as:

\begin{equation}
R_{\lambda,v+}=\sqrt{\frac{1}{N_+} \sum^{N_+} C_{i+}^{2}} ~~~~ R_{\lambda,v-}=\sqrt{ \frac{1}{N_-} \sum^{N_-} C_{i-}^{2}}
\end{equation}

\noindent where $C_{i+}$ is the raw correlation of one simulation in a specific wavelength bin $\lambda$ and line width bin $w$ (corresponding to the Gaussian line being placed at wavelength $\lambda$, with velocity width $v$) which is positive. The sum was then performed over all positive correlations with these Gaussian parameters. $N_{+}$ is the number of these positive $C_{i+}$ values. $C_{i-}$ and $N_{-}$ are identical variables but for negative raw correlations within the same parameter bin.

Once the normalization factors were calculated from the raw correlations of all 10000 simulated datasets, the raw correlations of the real dataset were reloaded into memory and normalized by the $R_{\lambda,v+}$ (positive raw correlations) and $R_{\lambda,v-}$ (negative raw correlations) factors. We repeated the same procedure for the simulated datasets, block by block.

\subsection{Collecting results}
\label{collection}

As the normalized cross-correlations were being saved, they were ordered by value within their wavelength/velocity width bins. The normalized cross-correlation value of the real dataset at each parameter bin was compared with these ordered lists and thus we obtained the p-value of each bin in the real dataset (i.e. what fraction of simulated datasets showed stronger correlation or anti-correlation compared with the real dataset). This value gives the single trial significance of each bin.

At the same time we saved the strongest correlation and anti-correlation from each of the simulated datasets. At this stage we combined the results from all individual spectral model files, obtaining the strongest correlations and anti-correlations for each simulation, taking into account all the spectral model parameters searched. Afterwards, these two extreme values from each simulation were ordered and compared with the real dataset. We thus obtained the true p-value of each cross-correlation in the real data - for each searched parameter bin in the real data we determined the fraction of simulated datasets showing a feature (anywhere within the searched parameter range) stronger than the real one. The true p-value (true significance) therefore takes fully into account the look-elsewhere effect.

Given that we obtained the various significances for each wavelength bin in the real dataset for each of the spectral model parameters (12 different velocity widths), this was a considerable amount of data which needed to be filtered. We filtered only the strongest spectral features, selecting any correlation peak with the true significance higher than 1$\sigma$ (true p-value lower than 33\%). In case a peak at a certain wavelength appeared in searches with different velocity widths, we chose just the velocity width with the highest normalized correlation.

Finally, we ran an automated routine which took the wavelengths and velocity widths of the selected peaks and fitted Gaussian lines with such properties to the source spectrum directly (in \textsc{spex}). From the direct fit we recovered the statistical fit improvement upon adding the extra Gaussian to the broadband continuum (the $\Delta$C-stat value), as well as the photon and energy fluxes of the added line, and we calculated its equivalent width.

\section{Statistics of each searched dataset, with details of individual objects}
\label{ulxexpostats}

Table \ref{ulxexpo} shows the clean RGS exposures and the net RGS counts for each approach of every object studied in this work. We also show the hardness ratio of each spectrum calculated from the broadband spectral model such as H/(S+H), where H is the luminosity in the $2-10$ keV energy band and S is the luminosity in the $0.3-2.0$ keV band. The hardness ratio is calculated from absorption-corrected luminosities. Finally, the table also shows the number of significant line detections in each of the sources studied (for each approach).

\begin{table*}
	\begin{center}
	\caption{Clean RGS exposures (per detector) and total RGS net counts (both detectors combined) for each approach on every object in the ULX sample. The exposure column lists two exposures (RGS 1/RGS 2) in cases where they differed significantly. The fourth column lists the hardness ratios of each spectrum determined from the broadband spectral continuum fits. The final three columns show the number of significantly detected line features - the total number and the number of emission and absorption features, respectively.} 
	\label{ulxexpo}
	\begin{tabular}{cccccccc}
		\hline
		Object name&Approach&Clean exposure&RGS net counts&Hardness ratio&Detected lines&Emission&Absorption\\
		&&(ks)&&H/(S+H)&&&\\
		\hline
		Circinus ULX-5&0701981001&48&1191&0.609&1&1&0\\
		Circinus ULX-5&0824450301&118&2685&0.675&3&2&1\\
		Holmberg II X-1&0200470101&56&12818&0.264&3&2&1\\
		Holmberg II X-1&Stack1&17&2253&0.412&4&1&3\\
		Holmberg II X-1&FullStack&137&22088&0.263&4&4&0\\
		Holmberg IX X-1&0200980101&96&7115&0.496&0&0&0\\
		Holmberg IX X-1&FullStack&181&19555&0.579&2&2&0\\
		IC 342 X-1&Stack1&87&1744&0.288&5&4&1\\
		M33 X-8&FullStack&26/33&10448&0.498&1&1&0\\
		NGC 1313 X-1&Stack1&287&14778&0.486&5&3&2\\
		NGC 1313 X-1&Stack2&406&29922&0.430&8&4&4\\
		NGC 1313 X-1&Stack3&304&12177&0.454&6&4&2\\
		NGC 1313 X-2&Stack1&76&1962&0.473&2&1&1\\
		NGC 1313 X-2&Stack2&102&986&0.372&3&1&2\\
		NGC 1313 X-2&FullStack&177&2962&0.478&2&1&1\\
		NGC 247 ULX&Stack1&706/718&16757&0.0088&8&5&3\\
		NGC 247 ULX&Stack2&706/718&16757&0.0074&6&3&3\\
		NGC 300 ULX-1&0791010101&134&7013&0.560&1&1&0\\
		NGC 300 ULX-1&0791010301&77&3954&0.535&3&2&1\\
		NGC 4190 ULX-1&FullStack&43&2585&0.569&0&0&0\\
		NGC 4559 X-7&0842340201&72&3793&0.382&6&3&3\\
		NGC 5204 X-1&FullStack&162&8975&0.371&3&3&0\\
		NGC 5408 X-1&Stack1&237&21294&0.160&5&3&2\\
		NGC 5408 X-1&Stack2&238&19261&0.179&3&1&2\\
		NGC 5408 X-1&FullStack&664&53986&0.160&7&4&3\\
		NGC 55 ULX&0655050101&120&6038&0.083&7&5&2\\
		NGC 55 ULX&0824570101&135&4677&0.112&4&3&1\\
		NGC 55 ULX&0864810101&130&6009&0.117&4&1&3\\
		NGC 55 ULX&FullStack&385&16713&0.074&7&5&2\\
		NGC 5643 X-1&0744050101&109&953&0.600&2&2&0\\
		NGC 6946 X-1&0691570101&110&2959&0.202&5&3&2\\
		NGC 7793 P13&Stack1&226&6945&0.697&2&2&0\\
		NGC 7793 P13&FullStack&385/389&11025&0.686&1&0&1\\
		RX J0209.6-7427&0854590501&12&117516&0.758&10&3&7\\
		SMC X-3&0793182901&32&43018&0.761&2&2&0\\
		\hline
	\end{tabular}
	\end{center}
\end{table*}

Below we list basic details of the individual objects studied in this work. We explain the approaches applied in their analysis as well as list the best-fitting broadband continua.

In general, we analysed all the individual \xmm\ observations which are of high enough data quality. Many sources have individual observations with quality lower than the 1000 count threshold, for which reason we also made stacks of all the available \xmm\ observations (called FullStack) for those objects. We did not create a FullStack if the individual observations were of very high quality (around 15000 counts of more). We also created smaller stacks of observations performed very close in time to each other, thus minimizing any long-term variations in the spectral lines. Finally, we did not create a FullStack spectrum if the increase in counts compared with the best dataset of the same object was less than 50 per cent since the statistics would just be dominated by the best dataset.

\subsection{Circinus ULX-5}

There are two \xmm\ observations of this ULX which are of high enough quality for our analysis \citep{Walton+13}. We analyse them individually. We do not attempt to stack them into a single dataset because the gained data quality would not be sufficient (<50 per cent gain compared with the best individual observation), especially considering that the observations occurred many years apart.

Applying the three-component spectral model, we found that the soft blackbody was not required for a reasonable fit. This could be due to high neutral absorption column, found to be in the range of  $(6.3-7.4)\times10^{21}$ \pcm. The best-fitting powerlaw slope was found to be 1.9 and 2.5. The hot blackbody temperature was $2.0-2.2$ keV.

\subsection{Holmberg II X-1}

Holmberg II X-1 is a very well studied, nearby and bright ULX, with a wealth of \xmm\ observations. We used three different approaches to study it: we searched the individual observation 0200470101 which is of high data quality \citep[first studied by][]{Goad+06}. Secondly, Stack1 consists of two observations taken close in time to each other. Finally, we also examined the stacked spectrum from all the \xmm\ observations. 

The best-fitting broadband continuum consists of a powerlaw with a slope of $\Gamma\sim2-2.5$ (depending on the dataset used), and two blackbodies with temperatures of $0.15-0.16$ keV and $1.0-1.9$ keV. The neutral absorption column density was $0.9-1.3\times10^{21}$ \pcm.

\subsection{Holmberg IX X-1}

Holmberg IX X-1 is another well studied ULX, spectrally much harder than Holmberg II X-1. We chose two approaches to study it, following \citet{Kosec+18a}. We analysed the long observation 0200980101 individually, and also stacked all the \xmm\ observations.

All three spectral components were required for a good fit. We found a powerlaw slope of $2.0-2.6$, and blackbody temperatures of $0.10-0.11$ keV and $2.9-3.7$ keV. The best-fitting neutral absorption column was $2.0-2.5 \times 10^{21}$ \pcm. 

\subsection{IC 342 X-1}

There were 2 \xmm\ observations well aligned for an RGS analysis of IC 342 X-1, and in both cases the source spectrum could be partially contaminated by another nearby X-ray source. As the observations individually were not of high enough quality, we stacked them into a single dataset.

The three-component broadband fit of this dataset resulted in a powerlaw slope of 2.8, blackbody temperatures of 0.13 keV and 3.6 keV and a neutral absorption column $1.0 \times 10^{22}$ \pcm.

\subsection{M33 X-8}

M33 X-8 is the nearest extragalactic ULX but also the least luminous ULX in our sample. We stacked all four well-aligned \xmm\ observations of the source into a single dataset which is of good enough quality for our analysis. 

Using the three-component spectral fit, we found a powerlaw slope of 1.9, and blackbody temperatures of 0.17 keV and 1.3 keV, obscured by a neutral column $1.2 \times 10^{21}$ \pcm.

\subsection{NGC 1313 X-1}

NGC 1313 X-1 is one of the best studied ULXs with many long \xmm\ observations. We split these observations into three stacks based on the NGC 1313 X-1 X-ray flux and spectral shape, following the strategy of \citet{Pinto+20}. We do not attempt to stack all of the individual observations into a single dataset as the three stacks each result in datasets with excellent data quality.

The three-component model was required for an acceptable fit. We find powerlaw slopes in the range of $2.0-2.4$, blackbody temperatures of $0.17-0.51$ keV and $2.5-3.4$ keV, and a neutral obscurer column of $2.2-2.5 \times 10^{21}$ \pcm.

\subsection{NGC 1313 X-2}

NGC 1313 X-2 has been extensively studied as part of the \xmm\ campaigns on the brighter ULX in its host galaxy, X-1. However, most of the observations are not aligned well for an RGS analysis. We stacked all the available early observations (carried out between 2003 and 2005) into a single dataset named Stack1. Stack2 consists of two observations 0764770101 and 0764770401 which were performed close to each other. Finally, we also stacked all the available data (FullStack).

The best-fitting three-component continuum resulted into powerlaw slopes of $2.0-2.3$, blackbody temperatures of $0.11-0.17$ keV and $1.9-2.4$ keV and neutral absorption columns of $2.4-3.4 \times 10^{21}$ \pcm.

\subsection{NGC 247 ULX}

NGC 247 ULX is a very soft ULX which has recently been covered by a deep \xmm\ campaign of 8 long observations. Considering that these observations occurred close to each other and that the flux of the ULX is quite low, we stacked all of these spectra into a single dataset.

Initially we fitted the standard three-component spectral model. We found that only a single blackbody model was required. The powerlaw slope was very steep with $\Gamma\sim4.2$, the blackbody temperature was 0.13 keV and the neutral absorber column density was $3.9 \times 10^{21}$ \pcm. This best-fitting continuum is analysed in the approach Stack1. However, upon inspection of the results, we found a very strong positive residual around 1 keV which strongly influenced the line search in this region. For this reason, we tried a second broadband continuum, including a broad Gaussian line in this region to remove the very large effect of the 1 keV residual and hopefully reveal weaker spectral residuals in this band. The addition of the 1 keV Gaussian was highly significant with \delcstat$\sim$100. The best fit produces a powerlaw with a slope of 4.4, a blackbody with temperature 0.12 keV, a Gaussian with energy 0.92 keV and width 0.17 keV, all obscured by a neutral column of $4.0\times 10^{21}$ \pcm.

\subsection{NGC 300 ULX-1}

NGC 300 ULX-1 is a ULX pulsar \citep{Carpano+18} with two long \xmm\ observations performed close to each other (0791010101 and 0791010301). We analysed these separately as \citet{Kosec+18b} showed a variation in the spectral residuals between the two observations as well as due to their individually high RGS counts.

All three spectral components were required for a good fit. We found a photon index of $\Gamma\sim3.0-3.3$, blackbody temperatures $0.25-0.30$ keV and 2.9 keV and a neutral absorption column of $1.3-1.6\times 10^{21}$ \pcm.

\subsection{NGC 4190 ULX-1}

There are four well-aligned \xmm\ observations of this source, however they are all individually of insufficient data quality, hence we stacked them into a single dataset.

The continuum parameters recovered were as follows: powerlaw slope equal to 1.1, blackbody temperatures 0.13 and 1.7 keV and a neutral absorber column $1.7\times 10^{21}$ \pcm.

\subsection{NGC 4559 X-7}

There is only one \xmm\ observation of sufficient quality of this object, which we analysed here. 

The best-fitting model required all three components and resulted in the following best-fitting parameters: powerlaw slope 2.1, blackbody temperatures 0.15 keV and 1.5 keV and a neutral absorption column $1.6 \times 10^{21}$ \pcm.

\subsection{NGC 5204 X-1}

NGC 5204 X-1 has been observed many times by \xmm\ but all the individual observations were rather short. We therefore stacked all of the individual observations into a single dataset, following \citet{Kosec+18a}. 

The best-fitting continuum components consisted of a powerlaw with a slope of 2.3, two blackbodies with temperatures of 0.2 keV and 2.0 keV and a neutral absorber with a column density of $8\times 10^{20}$ \pcm.

\subsection{NGC 5408 X-1}

NGC 5408 X-1 has been observed many times by \xmm\ with deep exposures. Two of these campaigns resulted in observations being performed close to each other. We therefore created two dataset stacks (Stack1 and Stack2) from these campaigns. Additionally, we also created a stack of all the datasets which further increased the data quality but especially also included two other long observations not present in Stack1 and Stack2. Unfortunately, the orientation of the observations is such that another X-ray source is near the RGS source extraction region, and thus the ULX spectra could be partially contaminated. However, the EPIC data of the contaminating source show it is fainter and its spectra are much cleaner (featureless) than those of the ULX.

The best-fitting spectral model was as follows: a powerlaw with a slope of $2.7-2.8$, two blackbodies with temperatures of 0.14 keV and $1.2-1.4$ keV, and a neutral absorber with a column of $1.0-1.2\times 10^{21}$ \pcm.

\subsection{NGC 55 ULX}

NGC 55 ULX is a well-studied soft ULX which has recently been covered by three deep \xmm\ observations. We were able to analyse all of them individually. We also decided to create a stacked spectrum (FullStack), which resulted in further increase in data quality to bring out any weaker spectral features that are not time variable.

All three spectral components were required for an acceptable fit. We found powerlaw slopes in the range of $2.6-3.5$, blackbody temperatures of $0.15-0.17$ keV and $0.64-0.70$ keV. The neutral absorption column was around $2.3-3.1 \times 10^{21}$ \pcm.

\subsection{NGC 5643 X-1}

We analysed the only long observation of this distant ULX which is right on the border of our data quality cut.

We found that all three components were required. The powerlaw slope found was 1.6, and the blackbody temperatures were 0.16 keV and 1.8 keV, obscured by a neutral column of $1.8  \times 10^{21}$ \pcm.

\subsection{NGC 6946 X-1}

There are several \xmm\ observations of this well-studied ULX. However, only one (0691570101) was found to be of high enough quality to analyse in this work. 

The best-fitting continuum had a powerlaw slope of 2.4, blackbody temperatures of 0.17 and 1.5 keV, and a neutral absorber column density of $2.9\times 10^{21}$ \pcm.

\subsection{NGC 7793 P13}

NGC 7793 P13 is an extremely well studied pulsating ULX which has been observed many times by \xmm. However, most of these observations were rather short and thus did not collect the required 1000 counts in the RGS detectors. Thus we had to resort to stacking multiple observations. We created a single stack (FullStack) containing all the \xmm\ observations of P13. Additionally, we also created another stack (Stack1) containing only the observations that occurred close to each other in 2017. 

All three components were necessary for a good spectral fit. The best-fitting powerlaw slope was $2.8-2.9$, and the blackbody temperatures were 0.22 and 3.5 keV, with a neutral absorption column density of $1.5\times 10^{21}$ \pcm.

\subsection{RX J0209.6-7427}

RX J0209.6-7427 is an accreting pulsar in the Small Magellanic Cloud (SMC) which has recently gone into a luminous outburst. Considering the mass of the neutron star is tightly constrained and the X-ray luminosity was around $10^{39}$ erg/s, we know that the accretion rate of the object was highly super-Eddington during this event and thus we included it in this study of supercritical accretion. We use the only \xmm\ observation of this object. Due to the low distance of the object and its luminosity this is a very high quality dataset.

Considering the object belongs to a different class, its broadband spectrum somewhat differs from the average ULX spectrum. Instead of the usual three-component model, we use a model composed of the following components. First, a Comptonisation component (\textsc{comt} in \textsc{spex}) with a plasma temperature of 3.1 keV, a seed photon temperature of 0.05 keV, and an optical density of $\tau\sim10$. A simple powerlaw is not a good description of the broadband $0.3-10$ keV spectrum due to its strong curvature towards the higher end of this energy range but a \textsc{comt} component results in a very good fit \citep[see e.g.][for its application in the study of the X-ray pulsar Hercules X-1]{Kosec+20}. Secondly, we use a soft blackbody with a temperature of 0.13 keV and finally a broad Gaussian feature with an energy of 0.98 keV and a width of 0.27 keV. Using these components, absorbed by a neutral column of $5.8\times 10^{20}$ \pcm, resulted in a reasonable phenomenological fit, particularly in the RGS energy band.

\subsection{SMC X-3}

SMC X-3 is another X-ray pulsar located in the SMC which is known to enter super-Eddington outbursts occasionally. \xmm\ observed it during the latest outburst in 2016, and we analysed the observation (0793182901) in this work.

We employed a similar model to RX J0209.6-7427. We recovered the Comptonisation component plasma temperature of 3.7 keV, a photon seed temperature of 0.13 keV, an optical depth of 8.9 and a soft blackbody temperature of 0.13 keV. A broad $\sim$1 keV Gaussian feature was not necessary to obtain a good fit in this source. The neutral absorber column density was $1.9\times 10^{21}$ \pcm.


\bsp	
\label{lastpage}
\end{document}